\documentclass[12pt,a4paper]{article}

\usepackage[utf8x]{inputenc}
\usepackage[T1]{fontenc}
\usepackage[english]{babel}

\usepackage[a4paper,top=2.5cm,bottom=2.5cm,left=2.5cm,right=2.5cm,marginparwidth=1.75cm]{geometry}

\usepackage{amssymb,amsmath}
\usepackage{mathrsfs}
\usepackage{graphicx}
\usepackage[colorlinks=true, allcolors=blue]{hyperref}
\usepackage{hyperref}
\usepackage{authblk}
\DeclareMathOperator{\Lagr}{\mathscr{L}}
\usepackage{tikz}
\usepackage{subfigure}
\usepackage{indentfirst} 
\usepackage{cite}
\usepackage{gensymb}
\usepackage{textcomp}

\def\cm{{\,\rm cm}}

\def\s{{\,\rm s}}

\def\GeV{{\,\rm GeV}}
\def\TeV{{\,\rm TeV}}
\def\pc{{\, \rm pc}}

\title{The DAMPE excess and gamma-ray constraints}
\author[1]{Konstantin Belotsky}
\author[1]{Airat Kamaletdinov}
\author[2]{Maxim Laletin}
\author[1]{Maxim Solovyov}
\affil[1]{National Research Nuclear University MEPhI (Moscow Engineering Physics Institute), 115409, Kashirskoe shosse 31, Moscow, Russia}
\affil[2]{Space sciences, Technologies and Astrophysics Research (STAR) Institute, Universit\'{e} de Li\`{e}ge, B\^{a}t B5A, Sart Tilman, 4000 Li\`{e}ge, Belgium}
\date{}

\begin{document}
\maketitle

\noindent\rule{\textwidth}{1pt}

\begin{abstract}
The direct measurements of the cosmic electron-positron spectrum around 1 TeV made by DAMPE have induced many theoretical speculations about possible excesses in the data above the standard astrophysical predictions that might have the dark matter (DM) origin. These attempts mainly fall into two categories: i) DM annihilations (or decays) in the Galactic halo producing the broad spectrum excess; ii) DM annihilations in the nearby compact subhalo producing the sharp peak at $1.4 \TeV$. We investigate the gamma-ray emission accompanying $e^+e^-$ production in DM annihilations, as well as various theoretical means to suppress the prompt radiation, such as specific interaction vertices or multi-cascade modes, and conclude that these attempts are in tension with various gamma-ray observations. We show that the DM explanations of the broad spectrum excess tend to contradict the diffuse isotropic gamma-ray background (IGRB), measured by Fermi-LAT, while the nearby subhalo scenario is constrained by nonobservation in the surveys, performed by Fermi-LAT, MAGIC and HESS. We also briefly review other types of gamma-ray constraints, which seem to rule out the DM interpretations of the DAMPE broad spectrum excess as well. 

\end{abstract}

\noindent\rule{\textwidth}{1pt}

\section{Introduction}

The collective effort to detect any non-gravitational signatures of dark matter (DM), including multiple forms of direct and indirect searches, seems to progressively plunge into despair lately. Any experimental hints, such as
the famous detection of the annually modulating signal by DAMA \cite{Bernabei:2000qi},
the diffuse gamma-ray excess observed by EGRET \cite{Hunger:1997we} or
the $3.5$ keV line in the spectrum of galaxy clusters \cite{Bulbul:2014sua},
instantly gain a lot of attention in the community and induce a big wave of theoretical works. However, this enthusiasm turns to skepticism as the spectrum of proposed DM explanations faces arising theoretical inconsistencies, constraints or receives no confirmation from the other experiments. Even if the observed phenomenon persists in the new experimental data, there is usually an explanation, which does not require new physics, and it gradually gets more and more supporters. 

The search for the traces of DM annihilations or decays in charged cosmic rays (CR) is not an exception. 
The fraction of positrons in the $e^+e^-$ flux above $10 \GeV$, measured by PAMELA, \cite{Adriani:2008zr},
provided for the first time a clear evidence that the secondary production of positrons can not account for the observed signal. 
This discovery was later confirmed by Fermi-LAT \cite{FermiLAT:2011ab} and AMS-02 \cite{Aguilar:2013qda}, 
and triggered an avalanche of theoretical works advocating the DM origin of the \textit{positron excess} (a few notable examples include \cite{Feldman:2008xs,Grajek:2008pg,Nardi:2008ix,ArkaniHamed:2008qn,Meade:2009iu}). However, already the very first indications of this phenomenon by HEAT \cite{Barwick:1997ig} initiated this flow of papers \cite{Baltz:2001ir,Kane:2001fz,Hooper:2004xn},
which just got exponentially bigger after PAMELA. Eventually, this wave 
essentially crashed against multiple constraints, mostly concerning the emission of gamma rays, associated with the production of high-energy positrons (see e.g. \cite{Cirelli:2012ut,Kopp:2013eka,Ibarra:2013cra,Belotsky:2016tja,Blanco:2018esa}; also see Section~\ref{Discussion}). The exact origin of the positron anomaly is yet to be determined, though it appears that more attention is drawn towards the surrounding pulsars as a source of the major part of these positrons. This hypothesis was introduced shortly after the PAMELA discovery, although it was known long before that the pulsars should produce high-energy electrons and positrons (see \cite{Hooper:2008kg,Yuksel:2008rf} and references therein).
The advantages of this hypothesis include the fact, that we actually know about the existence of pulsars, located quite close to the Solar system, and that there are observations of their radio and gamma-ray emission in the region above TeV, which are consistent\footnote{However, some studies reveal inconsistencies coming from the sub-TeV gamma-ray observations of an extended area around Geminga and Monogem \cite{Shao-Qiang:2018zla}.} with the production, acceleration and diffusion of a sufficient amount of positrons to explain the positron anomaly \cite{Manconi:2018azw,Hooper:2017gtd,Profumo:2018fmz,Fang:2018qco}. Though pulsars are distributed around us much less isotropically than DM, the predicted level of anisotropy in the $e^+e^-$ flux is compatible with the latest constraints \cite{Manconi:2018azw}.
The pulsar explanation does not invoke new fundamental physics, however the standard physics involved is rather complicated, so a number of free parameters has to be introduced in order to fit the observed signal. Thus, the predictive power of this model leaves much to be desired.

A new wave of theoretical papers was initiated by the recent measurements of the flux of cosmic electrons and positrons from tens of GeV to a few TeV, performed by the satellite-borne detector DAMPE \cite{Ambrosi:2017wek}. 
Their data indicates two interesting features: i) a ``knee'' in the spectrum around 1 TeV, which can also be interpreted as a \textit{broad spectrum excess} in the region from tens of GeV to 1 TeV; ii) an outlying point at $\approx 1400 \GeV$, which resembles a \textit{sharp peak} in the spectrum. The presence of the spectral break was soon confirmed by another satellite-borne facility CALET \cite{Adriani:2018ktz}, though the first hints of this break were observed by ground-based telescopes, such as HESS \cite{Aharonian:2009ah}, MAGIC \cite{BorlaTridon:2011dk} and VERITAS \cite{Staszak:2015kza} within the previous decade. However, no other experiment has confirmed the existence of the aforementioned sharp peak yet and there seem to be no indications of it in the available data\footnote{For instance, the search for lines around 1 TeV in the $e^+e^-$ data by CALET \cite{Asaoka:2019krt} and Fermi-LAT \cite{Mazziotta:2017ruy} yielded negative results.}. The global statistical significance of the peak in the DAMPE data is currently less than $3\sigma$ \cite{Fowlie:2017fya}.

Despite the fact that the positron excess and the broad spectrum excess in the total $e^+e^-$ flux measured by DAMPE are not actually related, basically the same gamma-ray constraints should apply to the DM models, which are supposed to explain this probable excess. These difficulties for the interpretation of the DAMPE data are actually covered in the literature \cite{Jin:2017qcv,Yuan:2017ysv,Wang:2018pcc,Li:2018abw}, 
though they still seem to be somewhat underrated \cite{Gao:2017pym,Feng:2019rgm,Liu:2019iik}. The attempts to attribute the sharp peak in the data to DM are immensely more popular (very few examples include \cite{Cao:2017rjr,Cao:2017ydw,Liu:2017obm,Ding:2017jdr,Li:2017tmd,Chen:2017tva,Okada:2017pgr}).
These models have to rely on the assumption that there is a DM clump very close to the Solar system (within a few hundred parsecs) in order to make the predicted peak sharp enough \cite{Jin:2017qcv}. Besides, it allows for lower velocity-averaged annihilation cross sections $\sim 10^{-26} \cm^2/\s$, which are not constrained by the aforementioned gamma-ray observations. On the other hand, this scenario seems problematic for other reasons that we are going to discuss further. 

The main goal of our paper is to show that all the proposed DM explanations of the DAMPE spectral features are either ruled out by a variety of gamma-ray measurements or, at least, seem to be in great tension with them. Especially, we concentrate on the constraint on the explanations of the broad spectrum excess coming from the Fermi-LAT measurements of the isotropic gamma-ray background (IGRB) \cite{Ackermann:2014usa}. This constraint is rather strict and also it is almost impossible to avoid from the model-building point of view, because it is directly related to the production of electrons and positrons in the Galaxy, which is always accompanied by radiation. We study a number of ``exotic'' or ``tuned'' model solutions to inhibit the high-energy electromagnetic radiation. As for the explanations of the sharp peak, we calculate the gamma-ray signal from a typical nearby DM clump and compare it to the sensitivities of different gamma-ray telescopes. 

The structure of the paper is as follows.
In Section \ref{IGRB_constr} we consider the isotropic gamma-ray constraints on the DAMPE broad spectrum excess 
for a generic leptophilic model and more exotic models with intermediate cascades.
In Section \ref{subhalo_constr} we
compare the gamma-ray flux from a nearby subhalo and the sensitivity of different gamma-ray observatories to point sources and extended sources.
In Section \ref{Discussion} we discuss other types of gamma-ray constraints and conclude about the problems of DM interpretations. Also, there are two Appendices:
in Appendix \ref{Lagrangian} we show in detail, why the production of final state radiation (FSR) does not depend on the type of the interaction vertex given 
the same two-body process cross section, and Appendix \ref{AppFigs} contains some extra figures related to the analysis considered in Section \ref{IGRB_constr}.

\section{Isotropic gamma-ray background constraints}
\label{IGRB_constr}
\subsection{Calculation methods}

We start with the description of the methods and calculations that we use throughout our analysis of the broad spectrum excess. Our methodology follows in general terms
the one which we employed in \cite{Belotsky:2016tja}.

The injection spectra of electrons, positrons and prompt gamma radiation
are obtained in different ways depending on the type of the annihilation mode as 
is clarified below. The total $e^+e^-$ fluxes to be observed at the Earth after their propagation in the Galaxy and gamma-ray fluxes originated from $e^+e^-$ interactions with the intergalactic medium mostly due to
Inverse Compton Scattering (ICS) and bremsstrahllung (bremss) are calculated using the GALPROP code\cite{GALPROP1}.
For the DM density distribution we choose the common NWF model \cite{Navarro:1996gj}
\begin{equation}
    \rho(r)=\frac{\rho_0}{\frac{r}{R_s}\left(1+\frac{r}{R_s}\right)^2} \, ,
    \label{NFW}
\end{equation}
where 
$\rho_0=0.25 \GeV \cm^{-3}$, which corresponds to the local DM density of
$0.4 \GeV \cm^{-3}$, and
$R_s=24$ kpc. As a matter of fact, the choice of the density profile does not affect the results noticeably (for example, see the corresponding J-factors in \cite{Cirelli:2010xx}).

The total gamma-ray flux from DM annihilation is the sum of several contributions, including prompt radiation, ICS and bremss. The first one may, in its turn, consist of $\pi^0$-decay products (for $\tau$-mode) and FSR, which is practically inevitable for any mode. The contribution of the prompt radiation to the diffuse gamma-ray flux, which we then compare to the Fermi-LAT measurements, is given by
\begin{equation}
    \Phi_{\text{prompt}}(E_\gamma)=\frac{dN_\gamma}{dE_\gamma}\, \langle\sigma v\rangle \times\frac{1}{\Delta\Omega}\int^{100\text{ kpc}}_0\int^{90^\circ}_{20^\circ}\int_0^{2\pi}\frac{1}{4\pi r^2}\left(\frac{\rho}{2 M_X}\right)^2r^2\cos(\theta)\, dr \, d\theta\,  d\phi\, .
    \label{Fpromt}
\end{equation}
Here $dN_\gamma/dE_\gamma$ is the gamma-ray spectrum per one act of annihilation,
which we calculate using analytical expressions or Monte-Carlo simulations depending on the model that we consider (see Sec.~\ref{generic_leptophilic} and \ref{cascade_models}),
$M_X$ is the mass of DM particle, $\langle\sigma v\rangle$ is the velocity-averaged cross section (we distinguish DM particles from antiparticles, so the total density is divided by 2). The solid angle $\Delta\Omega$ 
corresponds to the region of the Fermi-LAT analysis $(l\in[0;2\pi]$, $b\in
[20\degree;90\degree]$).

We 
take the most conservative Fermi-LAT IGRB data (model B) \cite{Ackermann:2014usa} and account for the contribution of the unresolved astrophysical gamma-ray sources, mainly AGNs, blazars and star forming galaxies.
Different estimations show that most of the IGRB, if not all of it (see e.g. \cite{Linden:2016fdd} and references therein), can be explained by various astrophysical sources and this fact implies strong constraints on the production of high-energy particles by DM annihilations/decays in the halo (see e.g. \cite{Blanco:2018esa}). In this work we use a rather moderate estimation of the unresolved astrophysical contribution to the IGRB which follows from the analysis done by Fermi-LAT \cite{DiMauro:2016cbj}. We treat this contribution as a background when we compare the predicted DM signal to the IGRB data. However, it is important to note that the main conclusion of our analysis virtually does not change even if the contribution of unresolved astrophysical sources is not taken into account.

We use the total $e^+e^-$ background from \cite{Niu:2017lts}, which was obtained as the best-fit background model for a variety of cosmic-ray data.
In fact, it is well known that the modeling of the electron-positron background is quite uncertain and one can try to improve the fit of the DAMPE data by adjusting CR model parameters or the contribution of astrophysical sources like supernovae remnants or pulsars.
But we do not try to analyze here, what is the most probable $e^+e^-$ background, predicted at these energies by some realistic astrophysical models, and how uncertain it is. Put simply, our goal can be formulated as follows: given that there is a \textit{clear} evidence for a broad spectrum excess over the expected background, we want to show that it cannot be due to DM annihilation. 

Usually, one fits only the $e^+e^-$ data and then checks whether the fit is constrained by the gamma-ray data. Thus, one can, in principle, miss a region in the parameter space, which is allowed by the gamma-ray data and also provides an acceptable, though not the best possible, fit of the electron data. For this reason we introduce the following statistical criterion\footnote{Similar statistical criteria applied to the fitting of cosmic electron and isotropic diffuse gamma-ray data were also used in \cite{Liu:2016ngs,Kalashev:2016xmy}.} to account for both sets of observables at the same time
\begin{gather}
    \chi^2=
    d^{-1}\left[\, \sum_{\substack{\rm DAMPE
    }}\frac{\left(\Delta \Phi_{e}\right)^2}{\sigma_e^2}+
    \sum_{\substack{\rm Fermi
    }} \frac{\left(\Delta\Phi_{\gamma}\right)^2}{\sigma_{\gamma}^2}\, H\left(\Delta\Phi_{\gamma}\right) \right].\label{chi2} 
\end{gather}
Here $\Delta\Phi_i\equiv\Phi_{i}^{\rm (th)}-\Phi_{i}^{\rm (obs)}$, 
$\Phi_{i}$ are the predicted (\textit{th}) and measured (\textit{obs}) fluxes for $i = e,\gamma$ denoting $e^+e^-$ or gamma points respectively, $\sigma_{i}$ denotes the corresponding experimental errors and $d$ denotes the number of statistical degrees of freedom, which includes all the relevant DAMPE and Fermi-LAT data points. 
The first sum in Eq.~\eqref{chi2} goes over the DAMPE data points (we consider the energy range from 20 to 1600 GeV) and the second sum goes over the Fermi-LAT data points.
Since we do not try to fit the gamma-ray data, but rather not to go over the experimental limits, the terms in the second sum are non-zero 
only when $\Phi_{\gamma}^{\rm (th)}>\Phi_{\gamma}^{\rm (obs)}$, which is ensured by the Heaviside step function $H$.

For our purposes, it is convenient also to express the statistical criterion in the following form $\chi^2=\chi^2_e+\chi^2_{\gamma}$. Note, that everywhere in the text we refer to the reduced values of the test statistic, i.e. divided by the corresponding number of degrees of freedom (as in Eq.~\eqref{chi2}). One has to keep in mind that this number can be different for each kind of test statistic that we refer to.
For example, when we give the values of $\chi^2_e$ it is implied that the Fermi-LAT data points are not included in the number of degrees of freedom. For the sake of brevity we do not specify this hereinafter.

Depending on the type of the DM model under consideration, we minimize $\chi^2$ over different sets of model parameters. We also distinguish two following $\chi^2$ minimization procedures. In the first case, which we call ``\textit{e}-fit'', we minimize only $\chi^2_e$, pursuing the best fit of $e^+e^-$ data only (as it is commonly done), and then we 
add $\chi^2_{\gamma}$ 
for the given parameter values. Thus, we basically find a local $\chi^2$ minimum in the parameter space of a model.  
In the second case, we minimize both $\chi^2_e$ and $\chi^2_{\gamma}$ simultaneously and obtain the global $\chi^2$ minimum in the parameter space. 
We refer to this minimization algorithm as ``combined fit''.
As is mentioned above, this way is more flexible and allows for lower values of the test statistic at the expense of the lower quality $e^+e^-$-data fit.
We assume that the model is ruled out as an explanation if the lowest possible value of $\chi^2$ exceeds unity. As a matter of fact, most of the cases that we consider display appreciably larger lowest-possible values of $\chi^2$, so there is practically no need to discuss the possibility of a false rejection.

We do not consider decaying DM in our analysis for the following reason.
Due to the diffusive and dissipative nature of electron galactic propagation, high-energy electrons and positrons are not likely to reach the Solar system from the distances above $\sim 1$ kpc. Thus, this local $e^+e^-$ production rate is defined by the DM density (or its square) averaged over the close vicinity of the Solar system and is ``fixed'' by the data, which these models aim to explain. The same combination of quantities determines the rate of gamma-ray production, which do not experience comparable energy losses and come to us from the whole DM halo. For annihilating DM this rate decreases as a square of DM density, while for decaying DM the decrease is less steep, so its contribution to IGRB should be larger. Basically the same argument applies also to the extragalactic gamma rays. Although we expect decaying DM models to give a larger contribution to IGRB, the difference between annihilating and decaying DM models is not supposed to be drastic.

\subsection{Generic leptophilic model}
\label{generic_leptophilic}

The most common and simple kind of DM models that can produce a substantial amount of high-energy electrons and positrons is the \textit{leptophilic} model. The name stresses the fact that these DM candidates couple mostly to SM leptons, so a typical annihilation process would be
\begin{equation}
XX \rightarrow l^+l^- \, .
\label{X2}
\end{equation}
We parameterize this generic model with a set of quantities, including the mass of DM particle $M_X$, the velocity-averaged annihilation cross section $\langle \sigma v \rangle$ and two values of branching ratios to electrons and muons (the third branching ratio to tau leptons is fixed as the sum of branching ratios is equal to unity). 
The injection spectra for $e^+e^-$ and prompt gamma rays from each mode are obtained\footnote{We simulate the following process $\tau^+\tau^-\rightarrow Z\rightarrow l^+l^-$. Initial state radiation is switched off. The choice of an initial state and a mediator hardly affects the spectra at the energies we are interested in.} with Pythia \cite{2008CoPhC.178..852S}.

It is well known that tau-mode yields much more gamma rays than the other modes, which is mainly due to $\pi^0$ production in $\tau$-decays. 
On the other hand, the presence of this mode appears to improve 
the fit of $e^+e^-$, because the electron injection spectrum for this mode is gently sloping. 
Nonetheless,
let us try to fit the DAMPE data with only electron and muon modes allowed and point out that even in this case, when there is no other contribution to prompt radiation except FSR, there is a contradiction with the Fermi-LAT data. 

\begin{figure}[t]
    \subfigure[]{
    \includegraphics[width=0.49\textwidth]{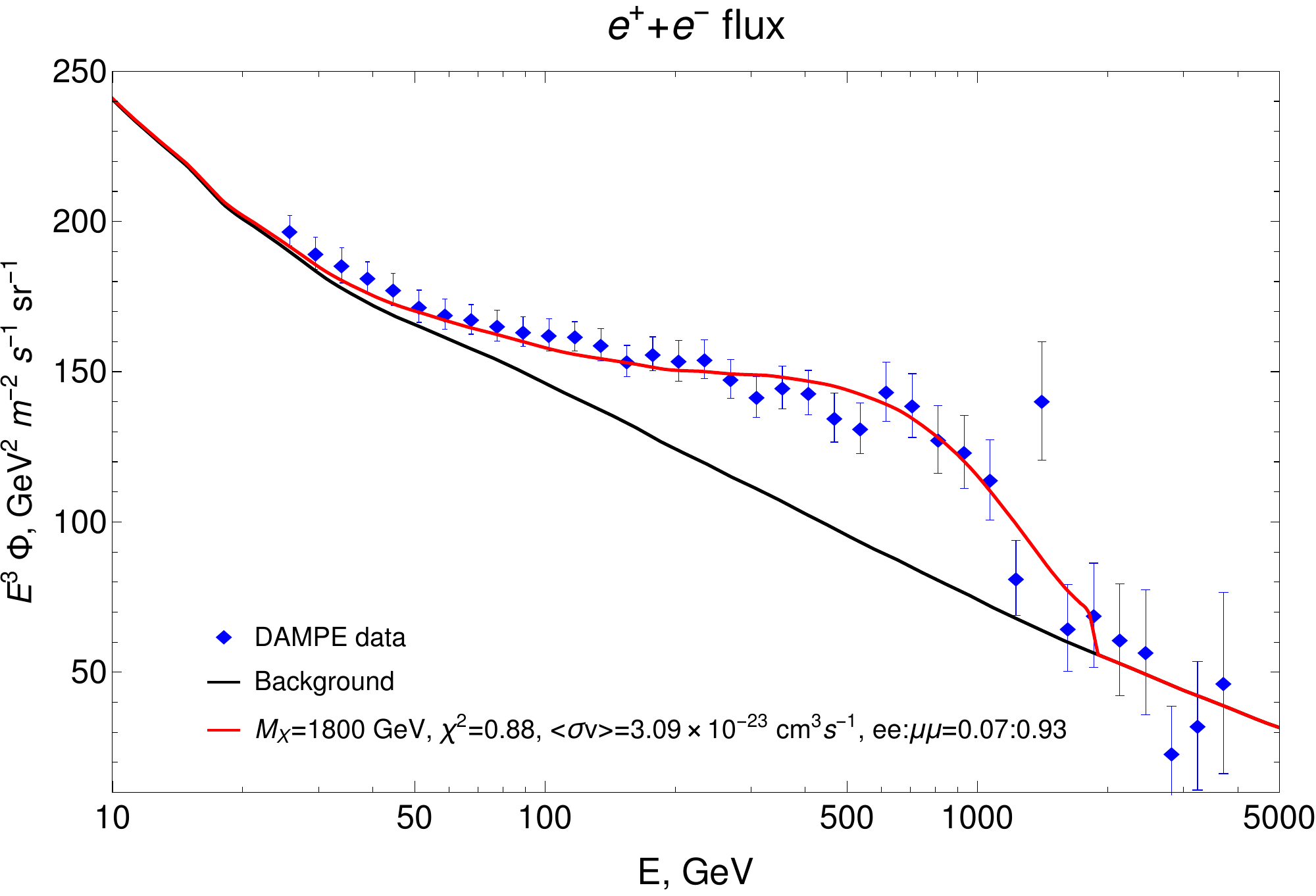}
    \label{2e_spectra_eemumu_e}}
    \subfigure[]{
    \includegraphics[width=0.50\textwidth]{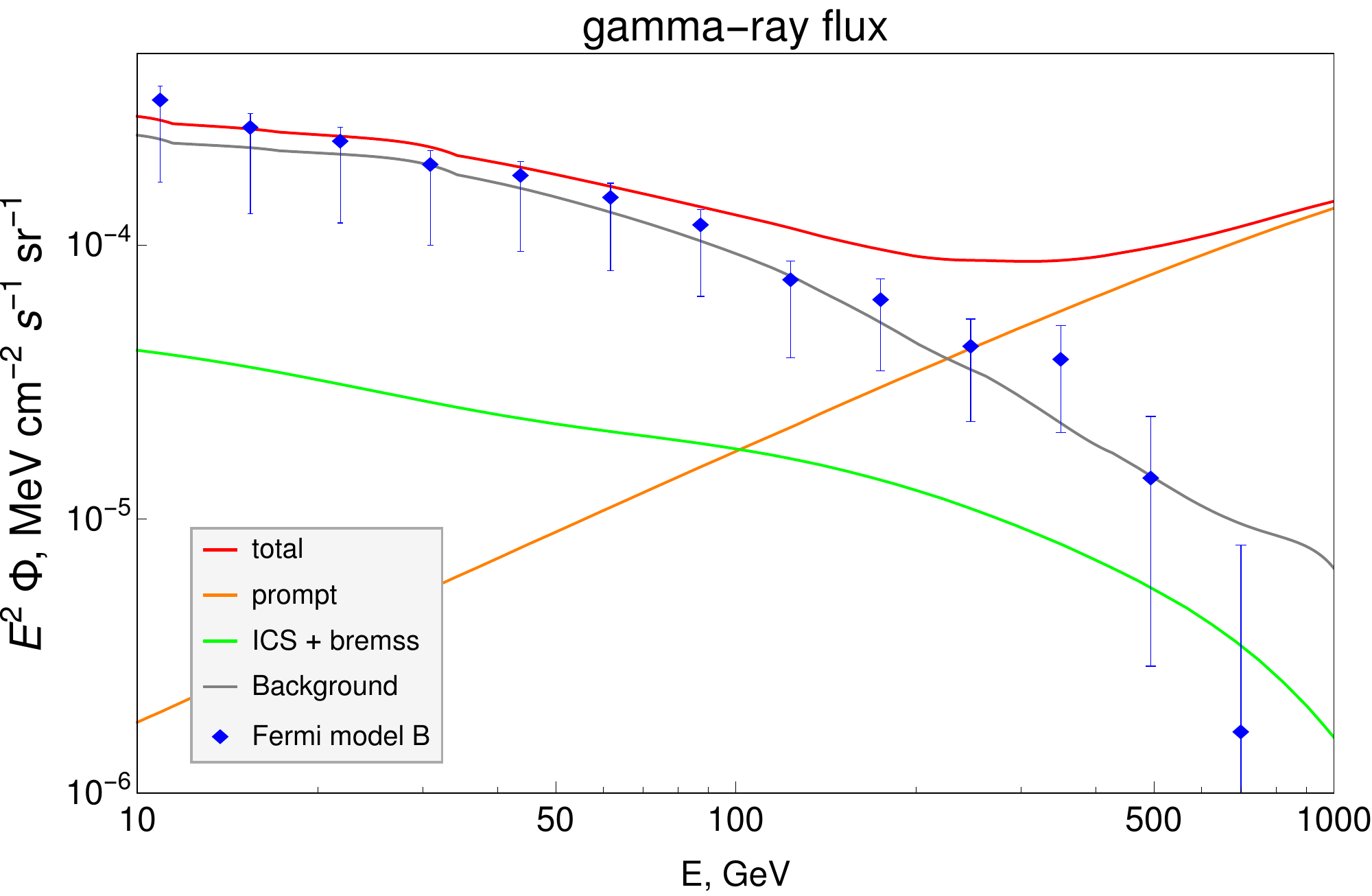}
    \label{2e_spectra_eemumu_gamma}}
    \caption{\textit{Left:} the total $e^+e^-$ flux from two-body annihilation to electrons and muons (excluding tau) in the best $e$-fit case (see text) compared to the DAMPE data. The parameter values are given in the plot legend.  \textit{Right:} the corresponding total diffuse gamma-ray flux (red curve) compared to the Fermi-LAT IGRB data. The other curves show different contributions (see the legend).}
    \label{2e_spectra_eemumu}
\end{figure}

\begin{table}[h]

\renewcommand{\arraystretch}{1.8}
\renewcommand{\tabcolsep}{0.4cm}
\begin{center}
\begin{tabular}{ |c|c|c|c| }
\hline
 & $XX\rightarrow e^+e^-,\mu^+\mu^-$ &  $XX\rightarrow e^+e^-,\mu^+\mu^-,\tau^+\tau^-$\\
\hline 
\textit{e}-fit & 4.6 (0.88) & 203 (0.53)  \\
\hline
Combined fit & 3.7 (2.0) & 3.8 (2.1) \\
\hline
\end{tabular}
\end{center}
\caption{\label{tab}
The lowest possible values of $\chi^2$ for different annihilation modes and minimization methods, described in the text. The corresponding values of $\chi^2_e$ are given in brackets.}
\label{table}
\end{table}

The results of this fit, \textit{viz.} $e$-fit as we define it, are shown in Fig.~\ref{2e_spectra_eemumu}. One clearly sees that while the $e^+e^-$ data can be nicely explained by electron and muon modes only, there is a significant excess of isotropic gamma rays at high energies. Quantitatively, this fact is demonstrated in the upper left cell of Table~\ref{table}. There one can also notice that the combined fit gives a lower value of $\chi^2$, as we argue above, though the main conclusion remains. Basically, the problem with FSR results from the shape of the energy spectrum: while the spectrum of IGRB is proportional to $E^{-2.3}$ with an exponential cut-off around 300 GeV, the spectrum of FSR scales as $E^{-1}$ (see Eq.~\ref{FSR1}) and has a sharp threshold far above 1000 GeV, which is required by the fit of $e^+e^-$ data. As we expect, if tau mode is included, $\chi^2_e$ decreases, but the corresponding $\chi^2$ skyrockets to an extreme value. The combined-fit minimization, in its turn, does not allow substantial values of the tau branching ratio, so the results do not actually differ. The plots illustrating the right column in Table~\ref{table} can be found in Fig.~\ref{2e_spectra_extra}. It is worth mentioning that, for simplicity, in this example we use a fixed value of $M_X = 1800 \GeV$ while optimizing the other parameters. However the variation of $M_X$ shows that our value is very close to the best-fit one.

Thus, we see that FSR causes a problem with the IGRB data, so one has to seek theoretical means to suppress it. Adjusting the physics of the dark sector in the leptophilic model without extending it, i.e. changing spins of annihilating particles and the mediator or changing the interaction vertices, cannot solve this issue (see Appendix~\ref{Lagrangian} for details). Another possibility to decrease FSR, which we know of, is to add intermediate steps to the production of $e^+e^-$ in the form of extra decaying particles in the dark sector. Such models of \textit{cascade annihilation} \cite{Mardon:2009rc} are considered below.

\subsection{Cascade models}
\label{cascade_models}

A model is called $n$-step cascade model if the final products of annihilation, e.g. electrons and positrons, are preceded by a consequent decay chain of $n$ types of particles. We concentrate only on 1-step and 2-step cascade annihilation models producing an even number of electron-positron pairs (and photons) in the final state, i.e. determined by the following processes
\begin{align}
    XX &\rightarrow a\bar a\rightarrow 2(e^+e^-),\label{X4}\\
    XX &\rightarrow b\bar b\rightarrow 2(a\bar a)\rightarrow 4(e^+e^-),\label{X8} 
\end{align}
where $a$ and $b$ are some intermediate-mass dark sector particles. We treat their masses as free parameters in addition to previously introduced $M_X$ and $\langle \sigma v \rangle$. We have chosen these two types of models, because they are rather generic, simple, and so they serve as nice demonstrative examples. As we are going to see,
these models can provide a good fit of the DAMPE data alone, so there is no need to introduce other lepton modes in the final decay. We do not consider cascade models with more steps for the reasons that we discuss further.

The key principle behind 
suppressing the FSR in cascade models is to decrease the energy available for radiation by electrons, which is $Q \approx m_a$, where by $m_a$ we always denote the mass of the last decaying particle in the cascade. It is, thus, reasonable to introduce new particles with the mass as close to the double mass of electron $m_e$ as possible. On the other hand, the spectrum of electrons gets deformed significantly, so one has to find a compromise between the two spectra. In case of 2-step cascade model \eqref{X8} the mass of the first particle in the cascade $m_b$ hardly affects both spectra, given that it is considerably larger than $m_a$, so we keep it fixed. Thus, as far as masses are concerned, we basically study the following parameter pattern $M_X > m_b \gg m_a \gtrsim m_e$.

We use analytical kinematic formulas for
the injection spectra of electrons and positrons in cascade processes. We treat $a$ and $b$ as free (on-shell) particles and neglect the effects of the matrix element, the interference between identical particles in the final state and the distortion of the spectrum due to radiation. 
The 1-step cascade annihilation 
gives a simple box-like spectrum of electrons and positrons.
For the 2-step cascade model \eqref{X8} $e^+e^-$ spectrum is derived as the convolution of the two box-like spectra of $a$ particles from $b$ decay and of electrons from $a$ decay at the given energy $E_a$

\begin{figure}[t]
    \subfigure[]{
    \includegraphics[width=0.5\textwidth]{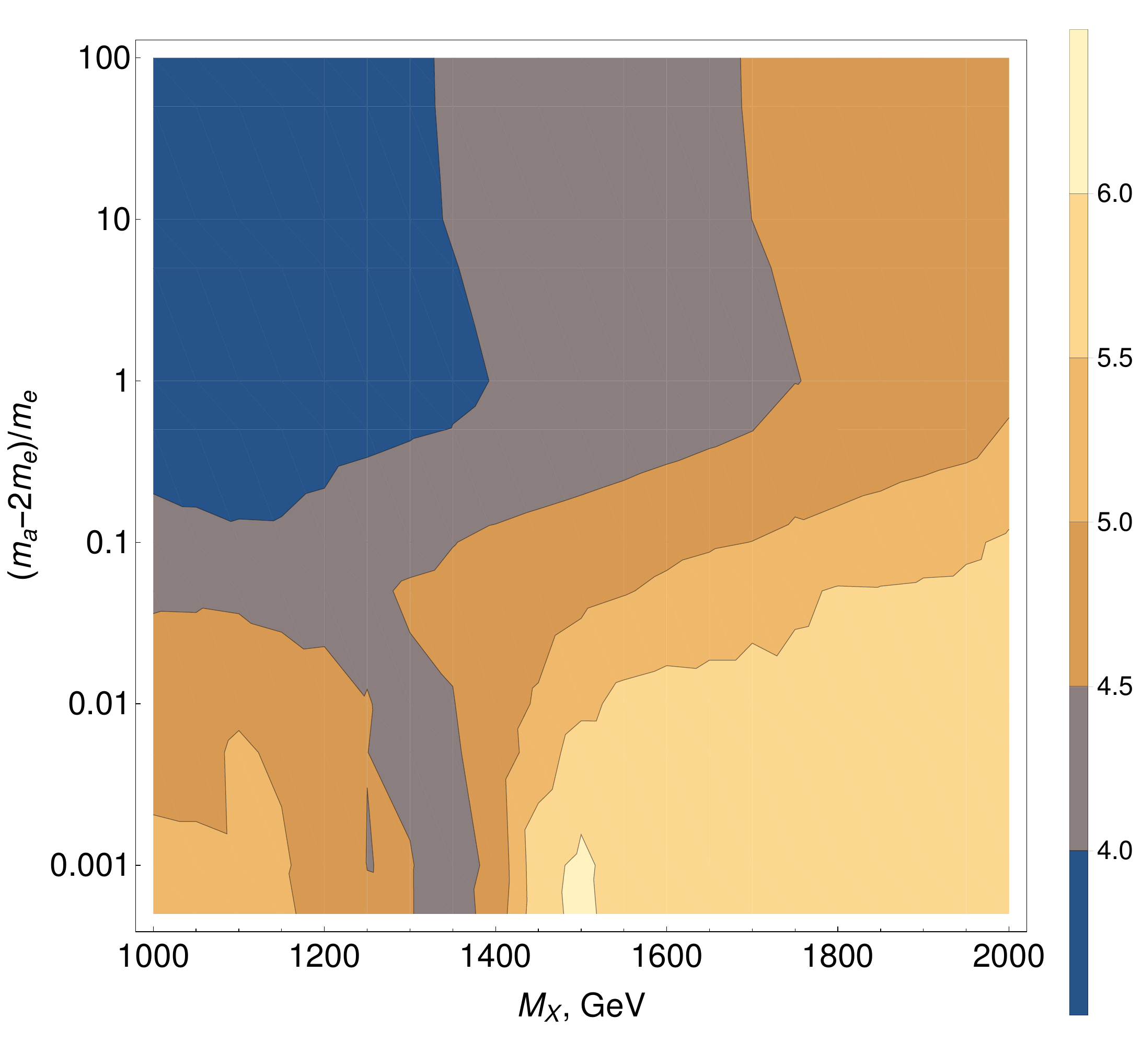}
    \label{4e_chi_main}}
    \subfigure[]{
    \includegraphics[width=0.5\textwidth]{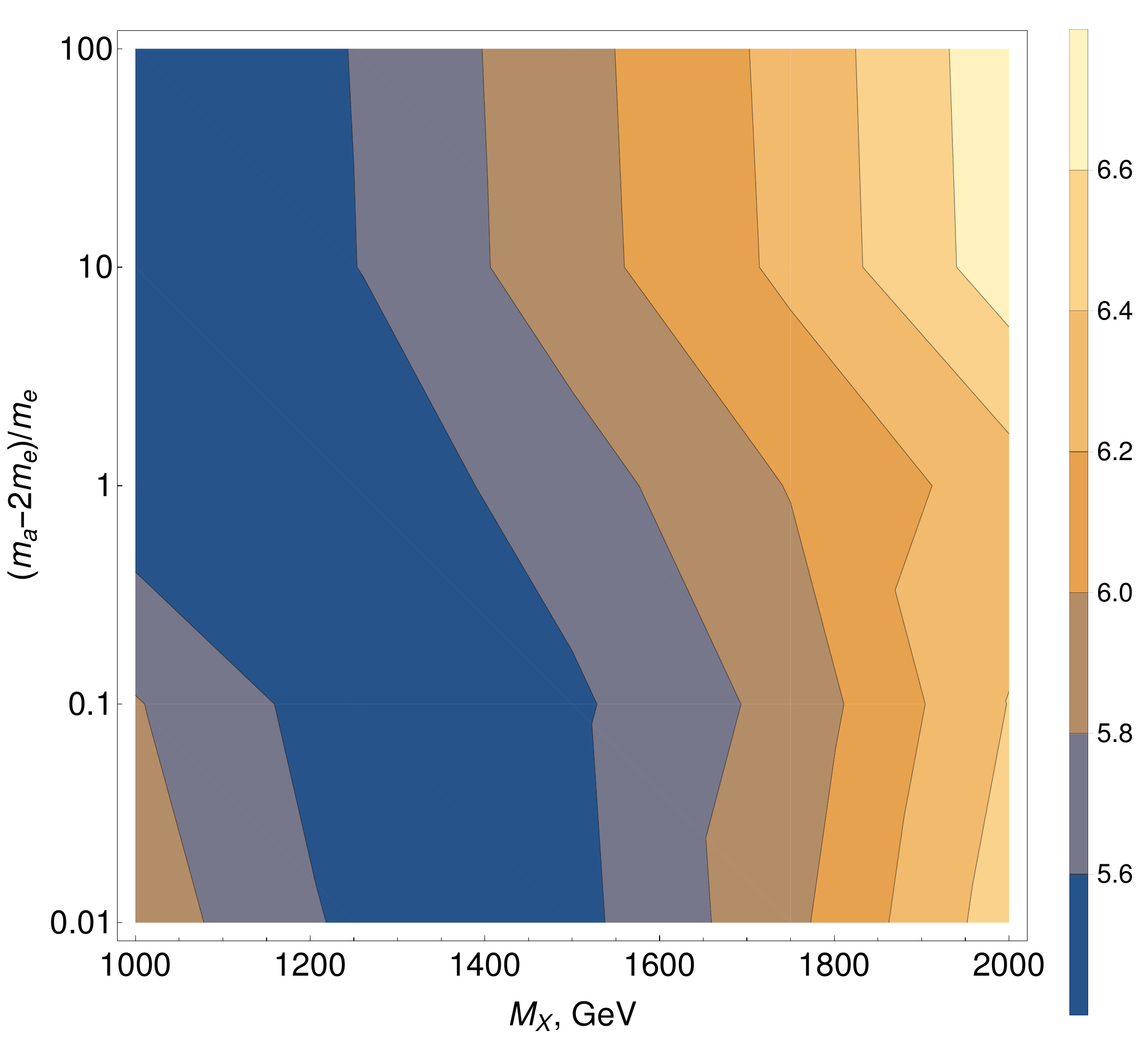}
    \label{8e_chi_main}}
    \vfill
    \caption{Contour plots for the best \textit{combined fit} values of $\chi^{2}$
    for 1-step \subref{4e_chi_main} and 2-step \subref{8e_chi_main} cascade models as a function of mass $M_X$ of the annihilating particle that initiates the cascade and
    mass $m_a$ of the last decaying particle in the cascade in units of electron mass $m_e$ minus 2. Note that here and in the similar figures below the color schemes in each plot are normalized differently. Plot \subref{4e_chi_main} has a better resolution than plot \subref{8e_chi_main} in terms of $M_X$ and $m_a$.
}
    \label{chi_plot_main}
\end{figure}

\begin{subequations}
\begin{equation}
  \frac{dN_e}{dE_e}=4\int^{M_X/2}_{2m_e}\frac{dw_{\, b\rightarrow aa}}{dE_a}\frac{dw_{\, a\rightarrow e^+e^-}}{dE_e}dE_a \, ,
  \label{fe2}
\end{equation}    
\begin{equation}
  \frac{dw_{\, b\rightarrow aa}}{dE_a}=\frac{2}{M_x}\Big|^{\frac{M_X}{2}}_{0} \, ,\label{btoa}
\end{equation}
\begin{equation}
    \frac{dw_{\, a\rightarrow e^+e^-}}{dE_e}=2\frac{m_a}{E_a p_e^*} \Big|^{E^+}_{E^-} \, .
    \label{wae}
\end{equation}
\end{subequations}
Here 
$E^{\pm}=(E_aE^*_e\pm p_ap^*_e)/m_a$, 
$m_e$, $p_e^*$ and $E_e^*$ are the electron mass, momentum and energy in the rest frame of $a$. The mass of $a$ in \eqref{btoa} is neglected (see above). 

The spectrum of FSR is calculated 
according to
\begin{subequations}
\begin{equation}
 \frac{dN_{\gamma}}{dE_{\gamma}}
=\int_E^{M_X}f_\gamma(E,E_0)\, \frac{dN_e}{dE_e}(E_0)\, dE_0\, ,
\label{FSR}
\end{equation}    
\begin{equation}
f_\gamma(E,E_0)= \frac{\alpha}{\pi E}\left(1+\left(1-\frac{E}{E_0}\right)^2\right)\left(\ln\left[\left(\frac{2E_0}{m_e}\right)^2\left(1-\frac{E}{E_0}\right)\right]-1\right).
\label{FSR1}
\end{equation}
\end{subequations}
Here $dN_e/dE_e$ is determined by Eq.~\eqref{fe2} for the 2-step cascade processes and by Eq.~\eqref{wae} for 1-step cascades respectively (multiplied by two).
Eq.~\eqref{FSR1} corresponds to the FSR energy spectrum emitted 
by an electron with energy $E_0$ \cite{Mardon:2009rc}.

In Fig.~\ref{chi_plot_main} we show the combined fit values of $\chi^2$ for 1- and 2-step cascade modes. 
The corresponding best-fit spectra of electrons and gamma rays for both $e$-fit and combined-fit minimizations can be found 
in Appendix \ref{AppFigs}.
both models fail to solve the problem with gamma rays (the minimal value of $\chi^2$ always significantly exceeds unity).  
Even though the yield of prompt radiation is actually reduced when $m_a\rightarrow 2m_e$, the $e^+e^-$ spectrum becomes distorted to an extent that it cannot properly fit the DAMPE data (see Fig.~\ref{cascades_chi_extra}). In fact, the problem of having a proper electron spectrum for $m_a \approx 2m_e$ outweighs the problem of having less gamma radiation in terms of the data description, especially in the case of 1-step cascade models, where the minimal $\chi^2$ is reached for $m_a > 2m_e$.

Extending the number of cascades 
does not improve the situation. For example, two-step cascades give rise to softer $e^+e^-$, which allows to fit the DAMPE data a little bit better. 
But this improvement is reached by introducing a larger value of $M_X$, which results in more high-energy gamma rays and therefore increases the minimal value of $\chi^2$.

\section{Gamma-ray constraints on the nearby subhalo scenario}
\label{subhalo_constr}

The constraints that we consider above, in principle, do not apply to the DM models, which aim to explain the sharp peak in the DAMPE data with a local source of high-energy electrons and positrons. The required value of the velocity-averaged annihilation cross section in this type of models is around $10^{-26} \cm^3 / \s$, so the annihilation signal from distant objects is rather weak to be constrained by the existing data (see Section \ref{Discussion}).
However, these attempts are still vulnerable to gamma-ray observations.
 
We start with one particular example, namely the model of the nearby ultra-compact micro halo \cite{Yang:2017cjm}. Indeed, an extremely dense compact subhalo in the $300$-pc vicinity of the Solar system can account for the sharp peak in the DAMPE data 
and escape IGRB constraints, if only $ \lesssim 10^{-5}$ of DM in the Universe form such dense structures \cite{Yang:2011eg}. 
The size of the dense core of this clump is, though, so small that it can be identified as a point source if it radiates sufficiently in gamma rays.

\begin{figure}[t]
    \centering
    \includegraphics[width=1\textwidth]{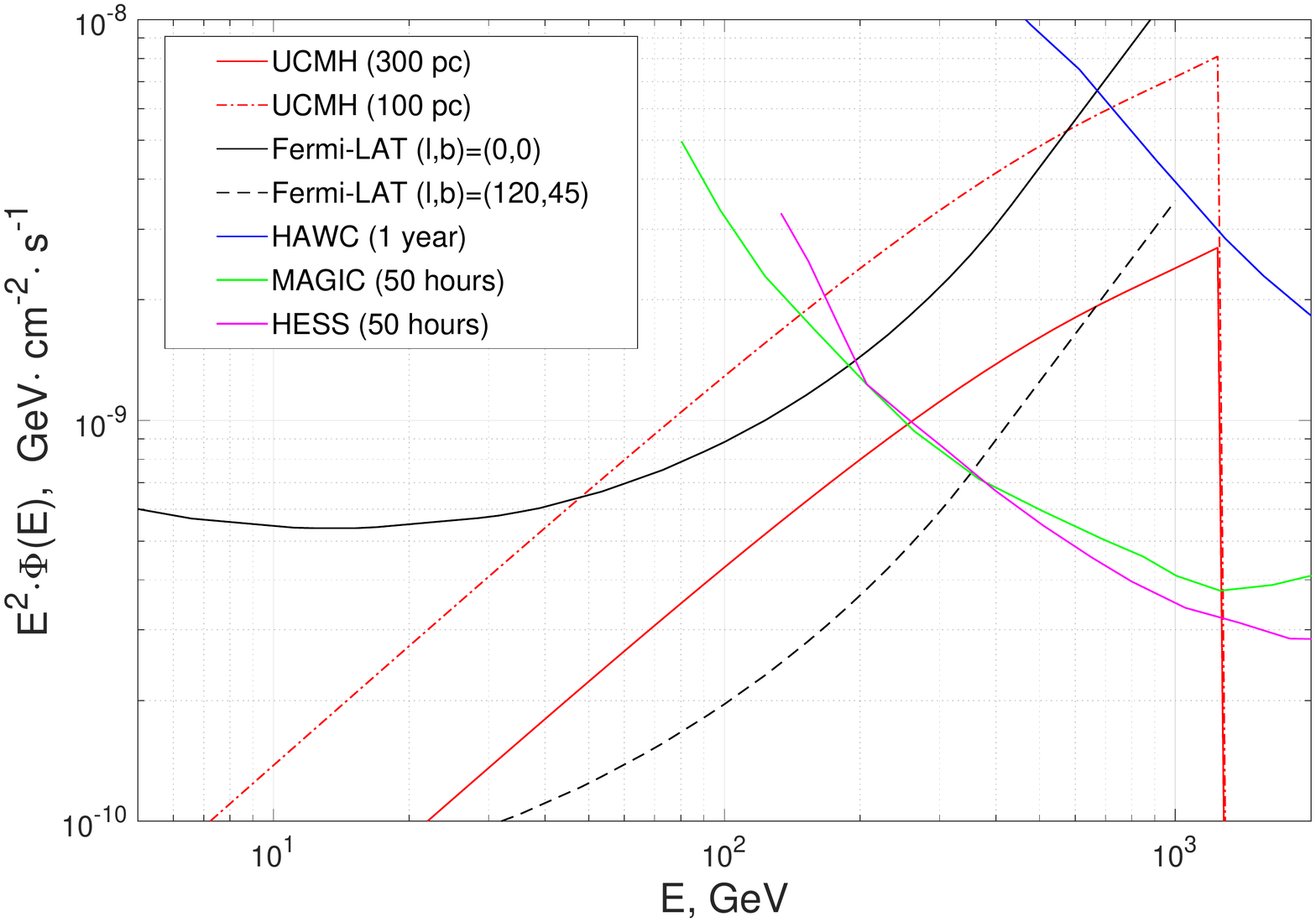}
    \caption{The gamma-ray fluxes from a nearby UCMH (red solid curve for the distance of $300 \pc$ and red dashed curve for $100 \pc$ respectively) compared to the point-source differential flux sensitivities of different gamma-ray telescopes, including Fermi-LAT (black solid and dashed curves) \cite{FermiLAT:performance}, HAWC (1 year, blue curve), MAGIC II (50 hours, green curve) and HESS (50 hours, magenta curve)~\cite{PointSourceSensitivities}. The black curves show the Fermi-LAT 10-year broadband sensitivities: the solid line corresponds to the minimal sensitivity in the direction of the GC and the dashed line corresponds to the maximal sensitivity in the direction of the Galactic periphery ($l = 120\degree, b = 45\degree$).}
    \label{UCMH_gamma_constr}
\end{figure}

One can estimate the injection power of the UCMH, needed to fit the peak in $e^+e^-$-flux at $E_0 =  1400 \GeV$, following the
monoenergetic\footnote{As a matter of fact, the typical value of the diffusion length for TeV electrons
is $\lambda \sim 100 \pc$. As it is similar to the distance from the source, the cooling effect can hardly be ignored. Taking this effect into account should decrease the flux of electrons with the maximal energy $E_0$ and, hence, increase the estimate of Eq.~\eqref{injection_power} within an order of magnitude. This makes our conclusions even stronger (see Fig.~\ref{UCMH_gamma_constr}).} electron approximation \cite{Yuan:2017ysv}

\begin{equation}
    \Dot{Q} = 3.12 \times 10^{35} \left(\frac{D}{100 \pc}\right) \frac{\GeV}{\s} \, ,
    \label{injection_power}
\end{equation}
where $D$ is the distance between the Earth and the source. The same value determines the gamma-ray emmissivity of the clump. The corresponding gamma-ray flux for the observer on Earth can be calculated as

\begin{equation}
    \Phi(E) = \frac{1}{4\pi}\frac{\Dot{Q}}{I \times 2E_0} \, f_\gamma(E,E_0) \int_{\Delta \Omega} d\Omega \int_{\rm l.o.s.} d\vec{s} \; \rho^2(\vec{s}) \, ,
\end{equation}
where $\rho(r)$ is the density profile of the clump, $I = \int dV \rho^2$, $f_\gamma(E,E_0)$ is the energy spectrum of FSR, 
and the integration goes along the line of sight in the direction of the clump within the angular resolution $\Delta \Omega$ for point sources. In fact, for an ultra-compact clump with a homogeneous and extremely dense core the formula above is very well approximated by a much simpler one 

\begin{equation}
    \Phi(E) = \frac{1}{4\pi D^2} \: \frac{\Dot{Q}}{2E_0} f_\gamma(E,E_0) \, .
\end{equation}
Note, that the injection power $\Dot{Q}$ is proportional to the first power of $D$, so the flux still depends on the distance to the source and is inversely proportional to it. For $e^+e^-$-annihilation channel the FSR spectrum is given by Eq.~\eqref{FSR1}.
The ICS contribution from a local DM clump is negligible (see also \cite{Jin:2017qcv,Fan:2017sor}).

The resulting gamma-ray fluxes from UCMH located at different distances from the Earth ($100 \pc$ and $300 \pc$) are presented in Fig.~\ref{UCMH_gamma_constr}. We compare them to the point-source differential flux sensitivities of the existing gamma-ray telescopes, such as Fermi-LAT, 
HAWC, 
MAGIC 
and HESS. 
Our results indicate that a nearby UCMH producing a sharp peak in the DAMPE data could have been detected by Fermi-LAT and also probed by MAGIC and HESS. Strictly speaking, this argument does not explicitly rule out local UCMHs as an explanation, but it undermines the credibility of this model and has to be taken into account. 

Let us now consider another benchmark subhalo model, which is a NFW clump \eqref{NFW} with an overdensity of $\mathcal{O}(10^3)$. For the clump located at the distance $D = 100$ pc we take $\rho_0 \approx 34 \GeV / \cm^3$ and $R_s = 65$ pc ($\rho_0 \approx 21 \GeV / \cm^3$ and $R_s \approx 200$ for $D = 300$ pc), which correspond to the mass and luminosity required to explain the sharp peak in the DAMPE data \cite{Yuan:2017ysv}. Such a clump is much more diffuse than the UCMH and can hardly be regarded as a gamma-ray point source. Even if we consider the gamma-ray signal from the portion of the clump that fits within the angular resolution of $0.1\degree$, it would be at least an order of magnitude lower than the Fermi-LAT constraints in Fig.~\ref{UCMH_gamma_constr}. However, it might be possible to detect this clump as an extended gamma-ray source. We do not perform here a detailed analysis of the available data, but estimate the detection capabilities. For an extended source of an angular radius of $0.5\degree$ the integrated flux sensitivity of Fermi-LAT, MAGIC and HESS is at the level of $\sim 10^{-12} \cm^{-2} \s^{-1}$ (in different energy intervals though)\footnote{To be more precise, our estimations give the following values: $2.0 \cdot 10^{-12} \cm^{-2} \s^{-1}$ for Fermi-LAT \cite{Funk:2012ca}, $3.4 \cdot 10^{-12} \cm^{-2} \s^{-1}$ for MAGIC (50 hours) \cite{Aleksic:2014lkm} and $4.3 \cdot 10^{-12} \cm^{-2} \s^{-1}$ for HESS (25 hours) \cite{HESS:performance}.}. Comparing this value with the predicted integrated fluxes from the considered NFW clumps, given in Table~\ref{extend_source_table}, we come to the conclusion that such an extended source is probably detectable by Fermi-LAT.

\begin{table}[h]
\renewcommand{\arraystretch}{1.8}
\renewcommand{\tabcolsep}{0.4cm}
    \centering
    \begin{tabular}{|c|c|c|}
    \hline
        & $100$ pc & $300$ pc \\ \hline
      100 MeV -- 1000 GeV & $2.5 \cdot 10^{-12}$ & $3.0 \cdot 10^{-12}$ \\ \hline
      200 GeV -- 1400 GeV & $3.8 \cdot 10^{-13}$ & $4.5 \cdot 10^{-13}$  \\
      \hline
    \end{tabular}
    \caption{Gamma-ray fluxes ($\cm^{-2} \s^{-1}$) from NFW subhalos within the region of $0.5\degree$ angular radius for different distances from the Earth (see the text above) and integrated over different energy intervals. The upper interval corresponds to the energy range of Fermi-LAT and the lower corresponds to MAGIC and HESS. The ICS contribution is neglected.}
    \label{extend_source_table}
\end{table}

It is worth noting that the detailed analysis done in \cite{Ghosh:2018bbe} essentially confirms our statement on the detectability by Fermi-LAT and, furthermore, claims that there is no significant evidence of the expected signal in the data. 

\section{Discussion and conclusion}
\label{Discussion}

We have studied the gamma-ray constraints on various DM related explanations of the possible excess of high-energy cosmic electrons and positrons, detected by DAMPE. We calculated the diffuse gamma-ray emission from DM annihilations in the Galaxy for a number of different DM models that are able to explain the broad excess in the electron-positron spectrum from $\sim 50 \GeV$ to $\sim 1 \TeV$ and compared it to the Fermi-LAT IGRB data, taking into account the contribution from the unresolved extragalactic astrophysical gamma-ray sources. We performed a statistical analysis over the allowed parameter range of these models to obtain the best \textit{possible} fit of the DAMPE data, which, at the same time, does not significantly exceed the IGRB limit. Three kinds of DM annihilation modes were considered: a) direct SM charged lepton pair production; b) electron-positron pair production through the 1-step cascade involving a new intermediate-mass decaying particle in the dark sector; c) electron-positron pair production through the 2-step cascade involving two different decaying particles in the dark sector. We have checked that the choice of the interaction vertex type in the Lagrangian cannot suppress FSR. Our analysis indicates that none of these models provide a satisfactory \textit{fit} of the two considered data sets, hence they are excluded as an interpretation of the broad DAMPE excess. 

The extragalactic contribution from annihilating DM substantially depends on the choice of DM distribution, so we did not include it in our calculations, though one should note, that our predictions are, in principle, underestimated. For this reason we actually did not consider decaying DM, because we know \textit{a priory} that it is a subject of even stronger constraints on the intensity of electron production, \textit{viz.} its lifetime. Indeed, as we argued above, the difference between gamma-ray emission from annihilation and decay is very subtle at high Galactic latitudes for our halo, but as DM distribution in the Universe is highly inhomogeneous, the averaged extragalactic diffuse gamma-ray emission is larger for decays in comparison with annihilation and this contribution at certain energy ranges can even exceed the Galactic one (see e.g. Fig.~6 in \cite{Ibarra:2013cra}). Note, that the IGRB constraint on the lifetime of DM decaying to SM leptons is $\tau \gtrsim 10^{28} \s$ for $M \sim 1 \TeV$ \cite{Blanco:2018esa}, whether the explanation of the DAMPE excess typically requires $\tau \sim 10^{26} \s$.

We do not consider $q\bar{q}$, $W^+W^-$ or $ZZ$ channels due to the fact that their spectra typically have lower $e^+e^-/\gamma$ ratios at high energies than lepton channels. This is why a proper fit of the high-energy part of the electron signal corresponds to a stronger tension with the IGRB limit. 
We limited our choice of possible steps in the annihilation cascade to 2.
Multi-cascade modes with a number of steps $> 2$ tend to yield a softer $e^+e^-$ spectrum for the given maximal energy threshold and a comparable spectrum of gamma rays, so they can hardly improve the combined fit. 
In addition, we regard the increase of the number of intermediate particles as a rather inadequate degree of fine-tuning. For the same reason, we did not consider models, which have different cross sections for each annihilation mode. 

In fact, the IGRB constraint is not the only argument against DM explanations of the broad DAMPE excess, related to gamma radiation. A few particular examples, which we find important to mention, are CMB constraints and the constraints coming from the gamma-ray observations of the GC and dwarf galaxies. 

\vspace{\baselineskip}
\noindent \textit{Cosmic microwave background.} \hspace{1 pt} If DM injects a sufficient amount of ionizing radiation through annihilations in the early Universe, it can alter the anisotropies of the CMB by modifying the last scattering surface. The detailed analysis performed in
\cite{Slatyer:2015jla} indicates that the cross section of $s$-wave annihilation, producing electrons and positrons, for DM particles with $M \sim 1 \TeV$ cannot exceed $\sim 10^{-24} \cm^3/\s$. In all of the cases that we have studied, the fit of the broad $e^+e^-$ excess requires annihilation cross sections of $\mathcal{O}(10^{-23} \div 10^{-24}) \cm^3/\s$.
In fact, this constraint can be circumvented by simply considering $p$-wave annihilation or decaying DM, to which the current CMB constraints in the corresponding range of parameters are not sensitive yet \cite{Diamanti:2013bia,Poulin:2016anj,Liu:2016cnk,Slatyer:2016qyl}.
Another common approach is to adjust a narrow-resonant annihilation cross section (see e.g. \cite{Bai:2017fav,Xiang:2017jou,Liu:2017obm}), which seem to require a larger degree of fine tuning.

\vspace{\baselineskip}
\noindent \textit{Dwarf galaxies.} Dwarf spheroidal galaxies (dSphs) provide an excellent aim for indirect DM searches. They typically have high dark-to-luminous mass ratios, own very little gamma-ray sources and most of them are located well above the Galactic plane, where the diffuse astrophysical foreground is low \cite{Strigari:2018utn}. If dSphs are dominated by the same kind of DM that creates the DAMPE excess, they can be used to impose constraints on the annihilation cross section. The combined analysis of 16 dSphs, performed by MAGIC and Fermi-LAT \cite{Ahnen:2016qkx}, seems to exclude $\langle \sigma v \rangle \gtrsim 10^{-24} \cm^3/\s$ for $M \sim 1 \TeV$ and for all the annihilation modes producing electrons and positrons. The limits are even stronger for some particular channels.

\vspace{\baselineskip}
\noindent \textit{The Galactic center.} The center of our Galaxy is probably the closest DM abundant region, where one would expect to observe a significant DM annihilation signal. Despite its relatively bright gamma-ray background and the uncertainty of the DM density distribution, some rather strong constraints on the annihilation cross section can be obtained. The analysis of 10-year data collected by HESS \cite{Abdallah:2016ygi} shows that even for a less cuspy Einasto profile the upper limit on the velocity-averaged annihilation cross section is as low as $\sim 10^{-25} \cm^3 /\s$ for $M \sim 1 \TeV$ for $\mu^+\mu^-$ or $W^+W^-$ channels and even lower for $\tau^+\tau^-$. 

\vspace{\baselineskip}
The possibility that the $1.4 \TeV$ peak in the DAMPE data originates from a nearby DM clump can not completely escape gamma-ray constraints. We showed that if an ultra-compact clump is located within $\sim 300 \pc$ from us and provides enough high-energy electrons and positrons to account for this peak, it radiates sufficiently in gamma rays to be detected as a point source by Fermi-LAT, MAGIC or HESS. A similar argument applies to a more diffuse clump with NFW density distribution, which can be detected as an extended source. If the clump is moved further away from the Sun, the peak becomes wider and does not fit the data so well \cite{Jin:2017qcv}. The nonobservation of such close-by clump cannot be considered as a solid refutation of this model, because this statement is possibly a subject of various subtleties and caveats. However, there are other arguments, which disfavor this explanation, mainly the extremely low probability that such a massive and compact DM subhalo would form close to us (see \cite{Jin:2017qcv,Genolini:2018jnu} and references therein). A nearby clump contributes to the dipole anisotropy of $e^+e^-$ flux at TeV energies at the level of a few percent \cite{Yuan:2017ysv}, which is comparable to the current Fermi-LAT limit of $\sim 10^{-2}$ at these energies \cite{Abdollahi:2017kyf}. Some studies \cite{Delos:2017thv,Gosenca:2017ybi} also indicate that the steep density profile of UCMH does not form in simulations under realistic conditions. The up-coming radio experiments will be able to test the hypothesis of a nearby compact clump in the near future \cite{Beck:2018aa}.

The presence of a broad excess or a spectral brake in the DAMPE data does not necessarily imply its DM origin. As we mentioned in the Introduction, pulsars are known to be powerful sources of high-energy electrons and positrons and can account for this effect (see e.g. \cite{Yuan:2017ysv,Wang:2018pcc}).
Another explanations, which do not rely on physics beyond SM, are based on the consequences of a very old local supernova event \cite{Fang:2017tvj,Kachelriess:2017yzq}. In fact, the explanation of the spectral knee might not require any specific additional sources at all, which was demonstrated in the stochastic CR sources approach \cite{Mertsch:2018bqd}. An extensive discussion of the general theoretical issues behind the spectral features of the DAMPE signal, related to the contribution of astrophysical sources and the specifics of the charged cosmic-ray propagation, can be found here \cite{Lipari:2019abu}.

As for the alternative DM hypotheses to account for the broad excess, that are not excluded by our analysis or any other arguments considered in this paper, one can think of some peculiar subcomponent of DM, which is capable of forming not-so-compact and sparse substructures. As an example, one can consider a thick dark matter disk emerging from the dissipative dynamics of some subdominant type of self-interacting DM, which also produces high-energy electrons and positrons through annihilations \cite[and references therein]{Belotsky:2016tja}. 
The existence of a thick dark disk is not excluded by the current observations of the dynamics of stars in the Milky Way \cite{Schutz:2017tfp} and if the dark disk is aligned with the baryonic disk the existing gamma-ray constraints on the production of a significant fraction of high-energy $e^+e^-$ can be also avoided \cite{Belotsky:2017wgi,Belotsky:2018vyt}. However, in the view of the uncertainties regarding the impact of known astrophysical sources on the total flux of high-energy electrons and positrons and the absence of a well established deficit in them, we find it rather prematurely to advocate this model as an explanation of the spectral features, observed by DAMPE.


\section*{Acknowledgements} 

We are grateful to R.~Budaev, whose computational code provided the basis of our work. We thank A.~Bhattacharya, J.~R.~Cudell, A.~A.~Kirillov and S.~G.~Rubin for useful comments and discussions. The work of MEPhI group was supported by the Ministry of Education and Science of the Russian Federation, MEPhI Academic Excellence Project (contract \textnumero~02.a03.21.0005, 27.08.2013). The work of K.B. is also funded by the Ministry of Education and Science of the Russia, Project \textnumero ~3.6760.2017/BY. The work of M.L. is supported by a FRIA grant (F.R.S-FNRS).


\bibliographystyle{JHEP}
\bibliography{References}


\appendix

\section{The impact of the interaction properties on FSR}
\label{Lagrangian}

\begin{figure}[h]
\centering
\subfigure[]{
\includegraphics[width=0.4\textwidth]{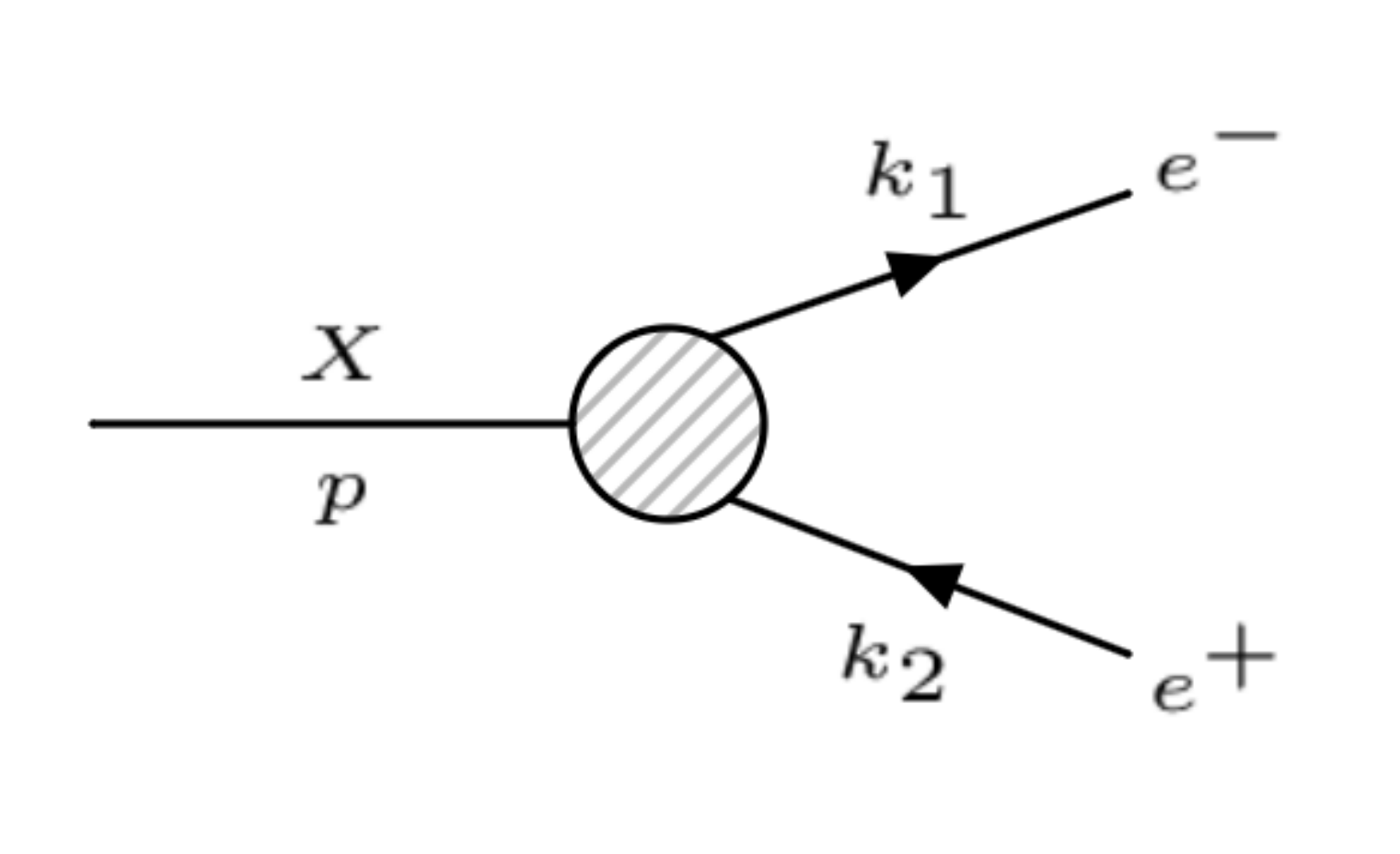}
\label{feyn1}}
\\
\subfigure[]{
\includegraphics[width=0.8\textwidth]{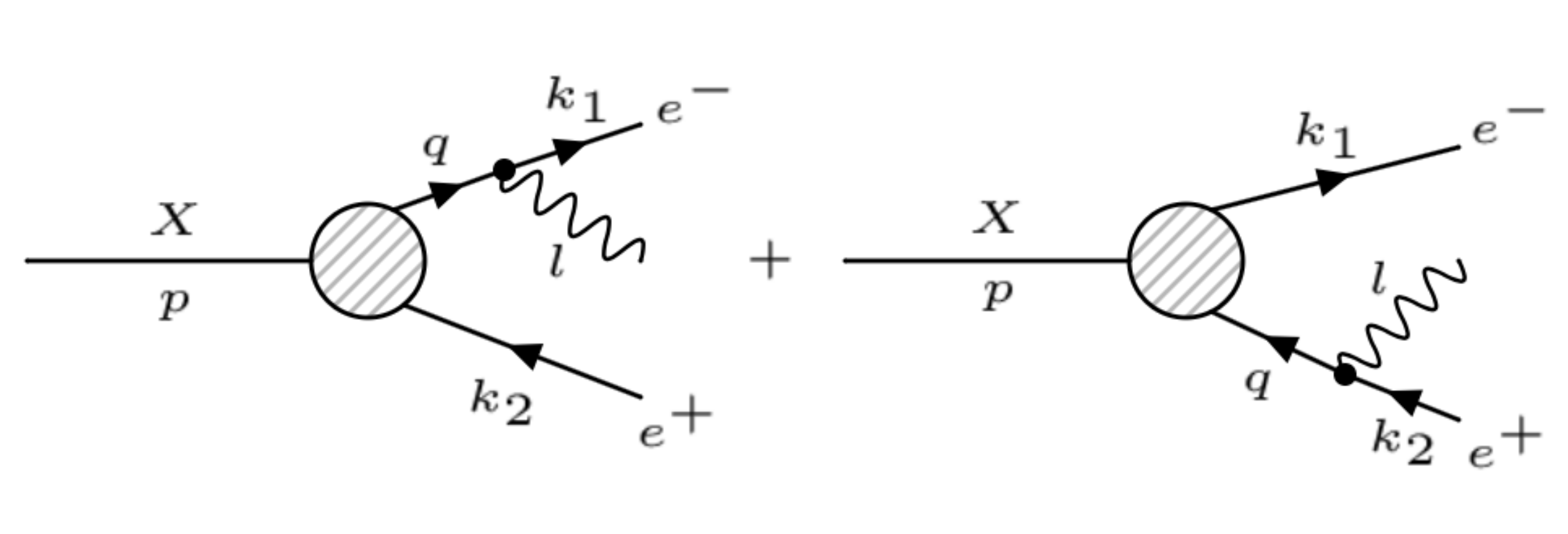}
\label{feyn2}
}
\caption{Dark sector particle $X$ decaying to $e^+e^-$-pair and emitting FSR.}
\label{Feynman}
\end{figure}

Here we show that the choice of some basic interaction vertices or spins of the particle decaying to electron and positron cannot suppress FSR. 
We consider the following two-body and three-body decay processes (Fig.~\ref{Feynman}).

The decaying particle $X$ can be a scalar, pseudoscalar, vector or axial-vector.
We parameterize the interaction Lagrangian in the following simple way for $X$ being scalar (pseudoscalar) and vector (axial-vector) respectively:
\begin{equation}
\Lagr_{scalar} = X\overline{\psi}(a+b\gamma^5)\psi \, ; 
\quad \Lagr_{vector} = \overline{\psi}\gamma^\mu (a+b\gamma^5)X_\mu \psi \, ,
\end{equation}
where $a$ and $b$ are some coupling parameters. To find a model that gives the least portion of FSR one has to vary these parameters, so that the following ratio of the decay widths of processes \ref{feyn1} and \ref{feyn2} is minimal

\begin{equation}
   R= \frac{\Gamma(X \rightarrow e^+ e^- \gamma)} 
   {\Gamma(X \rightarrow e^+ e^-)} \, . 
\end{equation}

The squared matrix elements of the considered two-body and three-body decay processes respectively for a scalar $X$ particle are:
\begin{equation}
|M|^2 = 4(a^2+b^2)(k_1 k_2) \, ; \quad |M|^2 = (a^2 + b^2)\lbrace \ldots \rbrace \, ,
\end{equation}
where the $\lbrace \ldots \rbrace$ part does not depend on $a$ and $b$ (see \cite{ICPPA:2018proceedings}).
Thus, the ratio $R$
is independent of $a$ and $b$,
hence, there is no suppression of radiation coming from the choice of the interaction vertex. The same conclusion holds if $X$ is a vector particle.

The reason why $R$ does not depend on those parameters is that the parity of permutations of vertex operators in $M M^\dagger$ does not change if a photon is introduced. In fact, this adds an even number of $\gamma^{\mu}$ from the photon vertex and the electron propagator ${(\hat{q}+m )/ (q^2 - m^2)}$. The term with $m$ in $M$ is multiplied by the same term in $M^\dagger$, so the parity of $\gamma^{\mu}$ matrices between the operators $(a+b \gamma^5)$ from $M$ and $M^\dagger$ is the same for 2- and 3-body decays.

\begin{equation}
|M|_{2-body}^2 \sim \text{Tr}\left(\hat{k_1}(a + b \gamma^5) \hat{k_2} (a - b \gamma^5)\right)
\end{equation}
\begin{equation}
\begin{gathered}
|M|_{3-body}^2 \sim \text{Tr}(...(a + b \gamma^5) \hat{k_2} (a - b \gamma^5) ...) 
+ \text{Tr}(...(a + b \gamma^5) \hat{k_2} \gamma_{\mu} \hat{q} (a - b \gamma^5) ) \\
+ \text{Tr}(...(a + b \gamma^5) \hat{q} \gamma^{\mu} \hat{k_2} \gamma_{\mu} \hat{q} (a - b \gamma^5)) + \text{Tr}( ... (a + b \gamma^5) \hat{q} \gamma^{\mu} \hat{k_2} (a - b \gamma^5) ... )
\end{gathered}
\end{equation}
%
%
The sum of $2 a b\cdot \text{Tr}(\gamma^{\mu_1} ... \gamma^{\mu_n} \gamma^5)$ terms, which arise from $(a + b \gamma^5)^2 = (a^2 + b^2) + 2 a b \gamma^5 $, cancels out. 

\newpage
\section{Supplementary figures}
\label{AppFigs}

\begin{figure}[h]
    \subfigure[]{
    \includegraphics[width=0.49\textwidth]{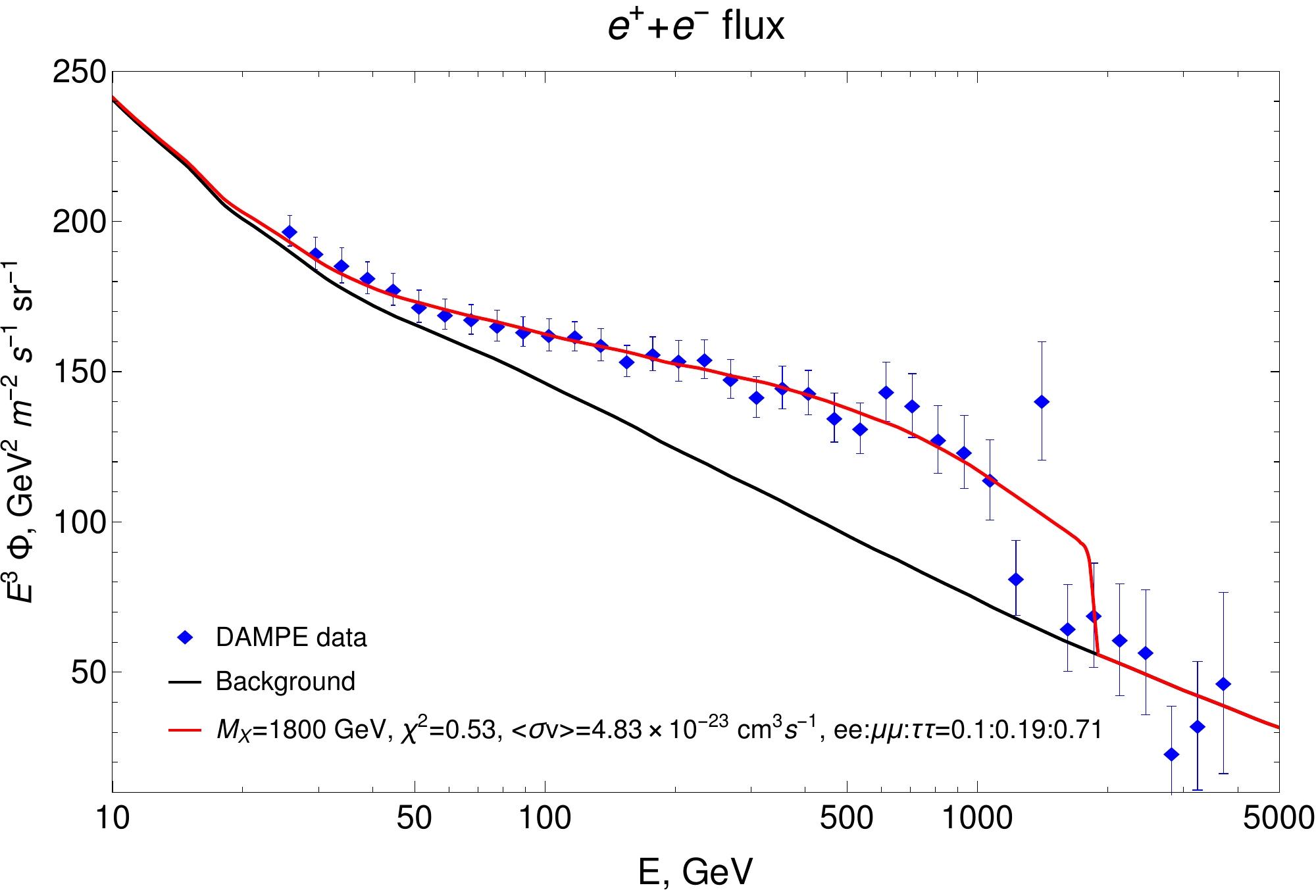}
    \includegraphics[width=0.5\textwidth]{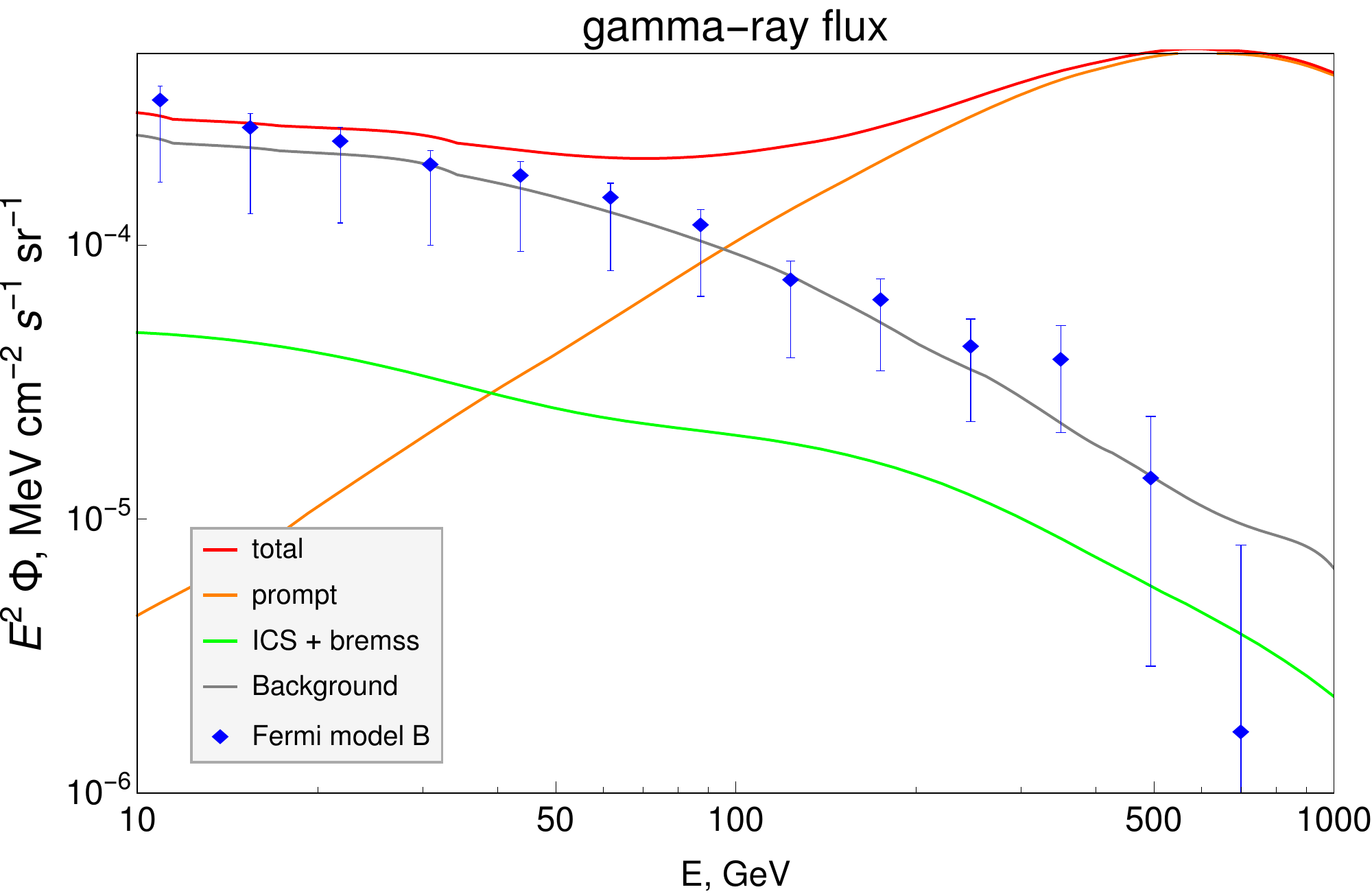}
    \label{2e_spectra_emutau_posifit}}
    \subfigure[]{
    \includegraphics[width=0.49\textwidth]{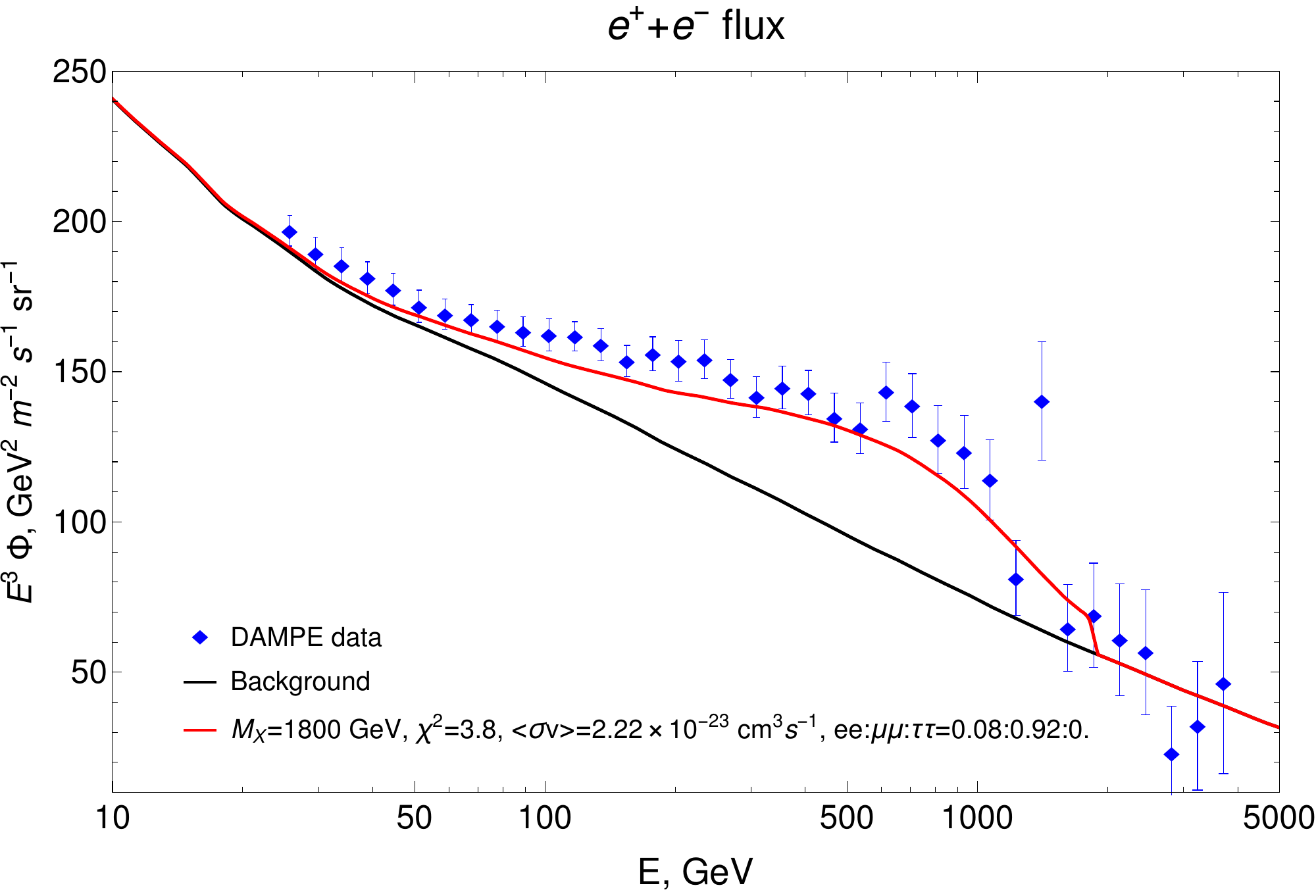}
    \includegraphics[width=0.5\textwidth]{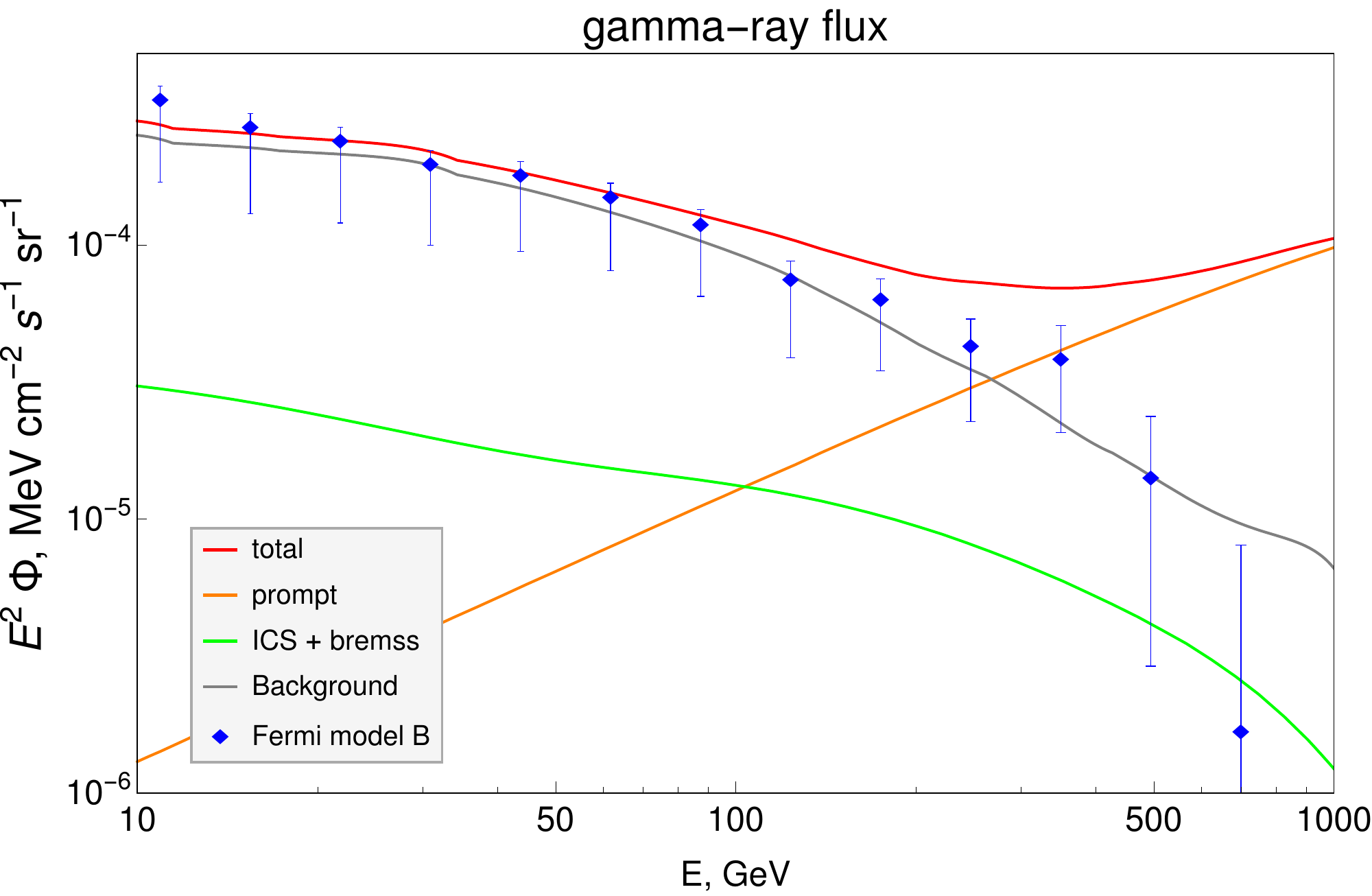}
    \label{2e_spectra_all}}
    \caption{Same as Fig.~\ref{2e_spectra_eemumu}, but for all leptonic modes, including $\tau^+\tau^-$. Plots \subref{2e_spectra_emutau_posifit} correspond to the best $e$-fit and plots \subref{2e_spectra_all} correspond to the best combined fit. 
    Note, that the combined fit favours the absence of the $\tau$-mode, so this case is similar to the one we show in Fig.~\ref{2e_spectra_eemumu}.}
    \label{2e_spectra_extra}
\end{figure}

\begin{figure}[p]
    \subfigure[]{
    \includegraphics[width=0.5\textwidth]{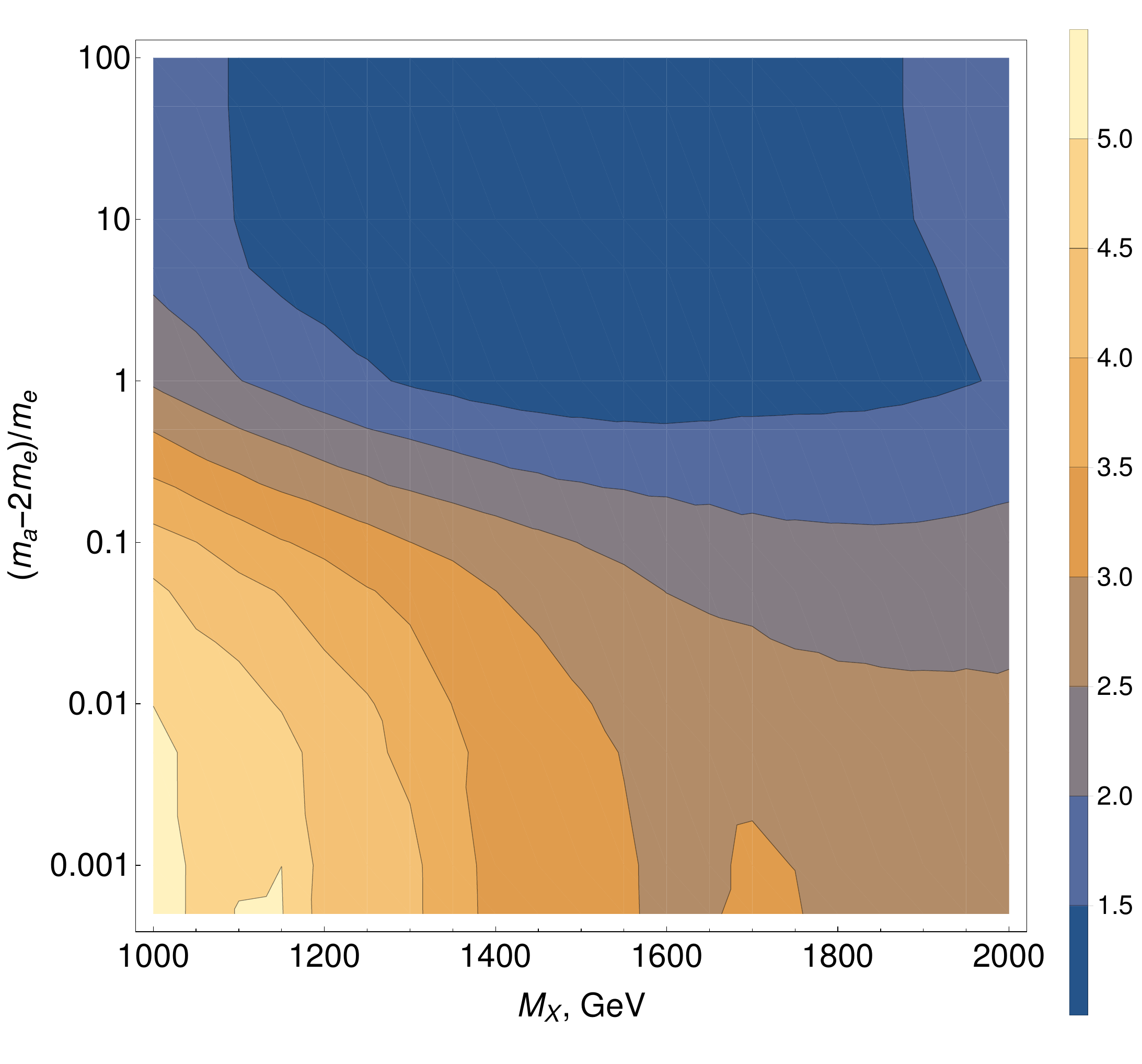}
    \includegraphics[width=0.5\textwidth]{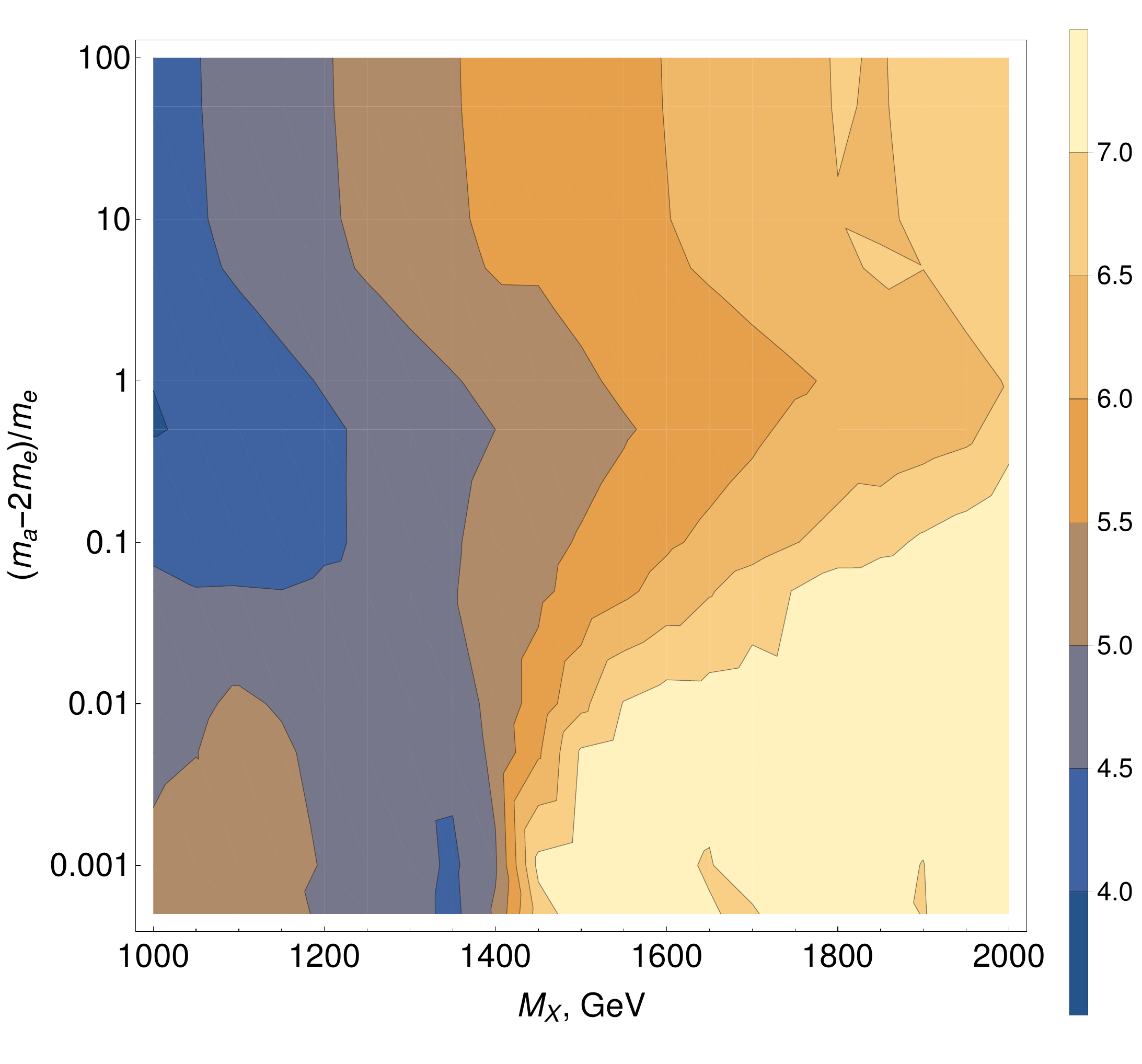}
    \label{4e_chi_posi}}
    \subfigure[]{
    \includegraphics[width=0.5\textwidth]{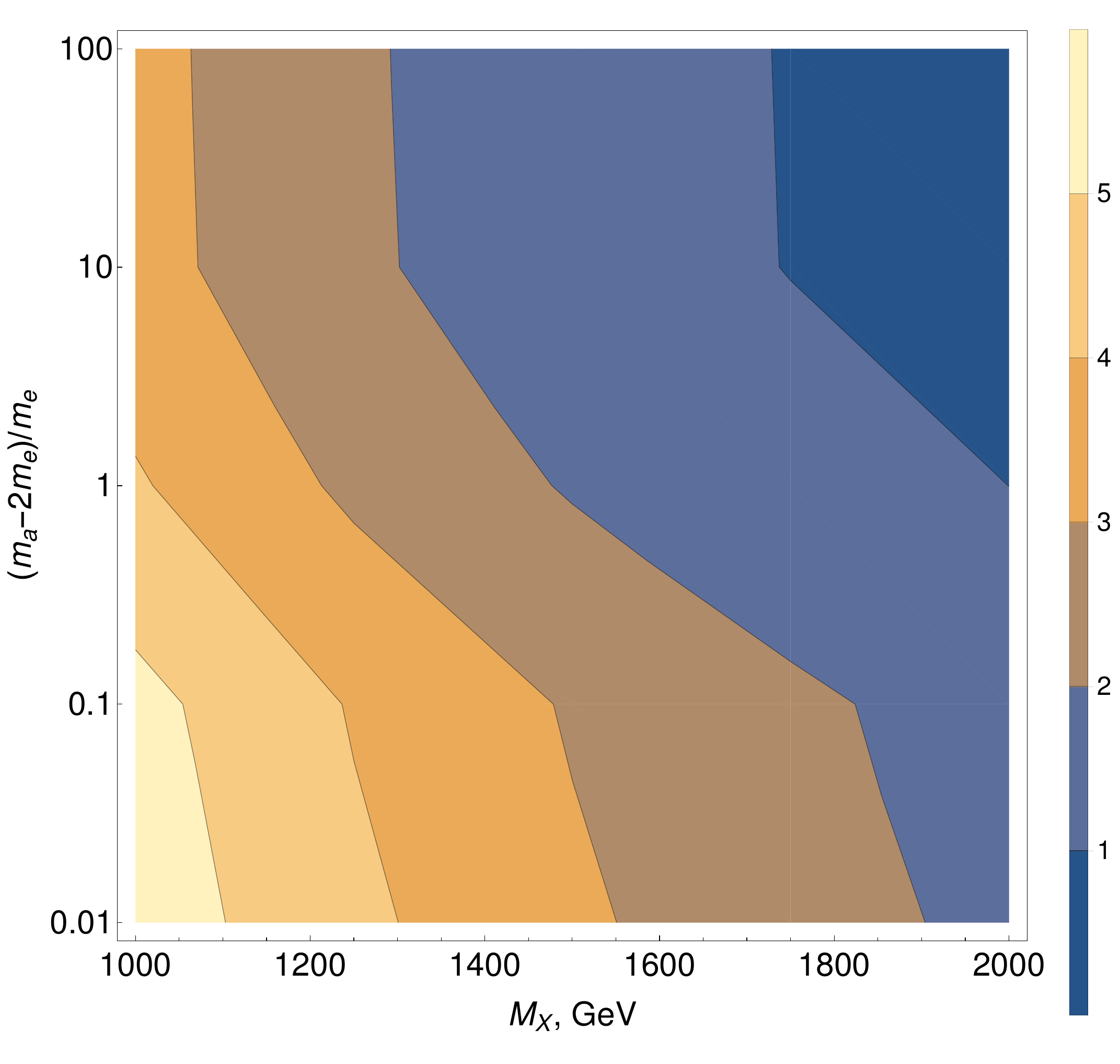}
    \includegraphics[width=0.5\textwidth]{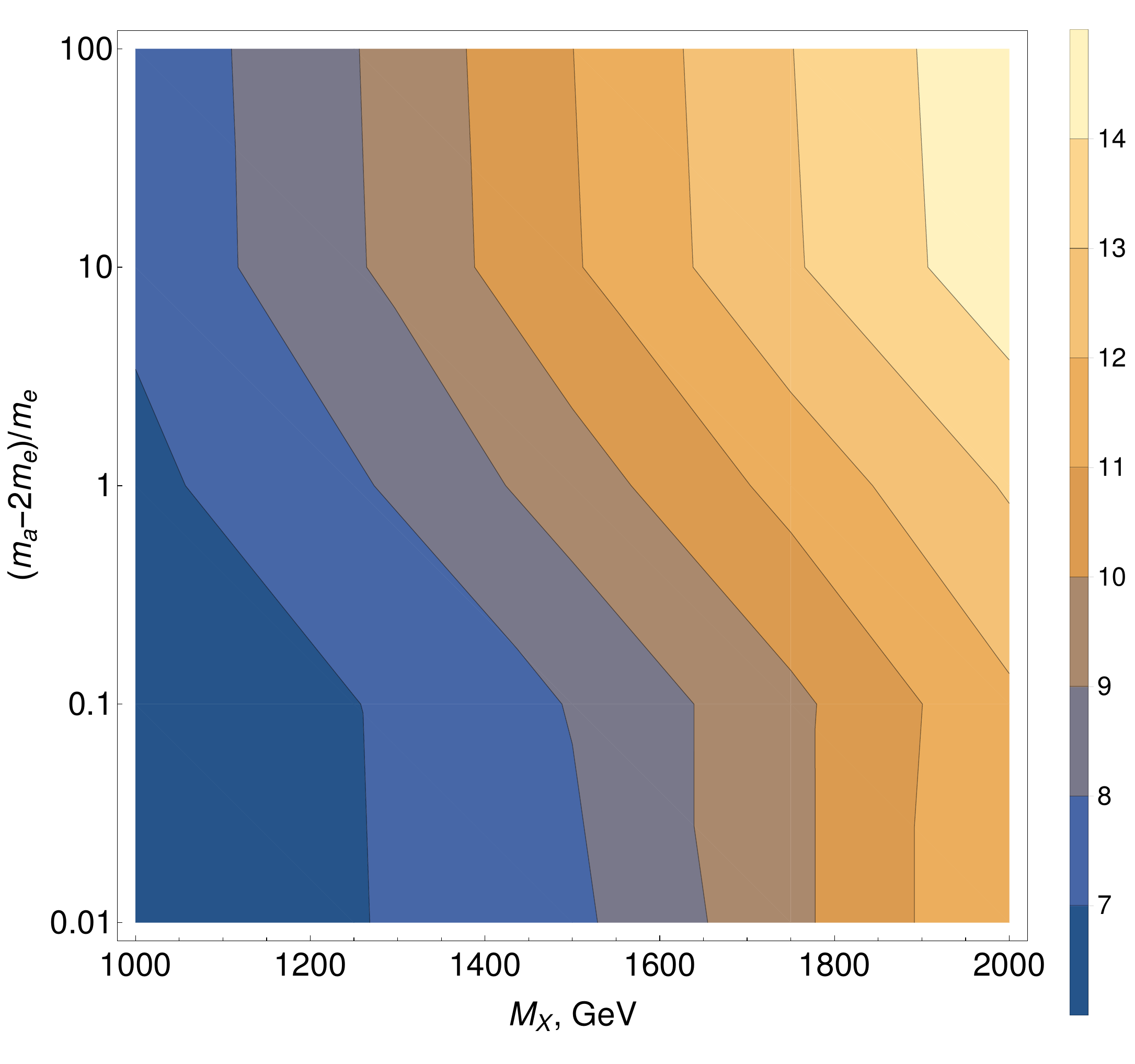}
    \label{8e_chi_posi}}
    \caption{
    Same as Fig.\ref{chi_plot_main},
    but the plots in the left column correspond to the best fit of the DAMPE data \textit{alone} without accounting for the IGRB constraint and the plots in the right column correspond to the best $e$-fit.}
 \label{cascades_chi_extra}
\end{figure}


\begin{figure}[p]
    \subfigure[]{
    \includegraphics[width=0.49\textwidth]{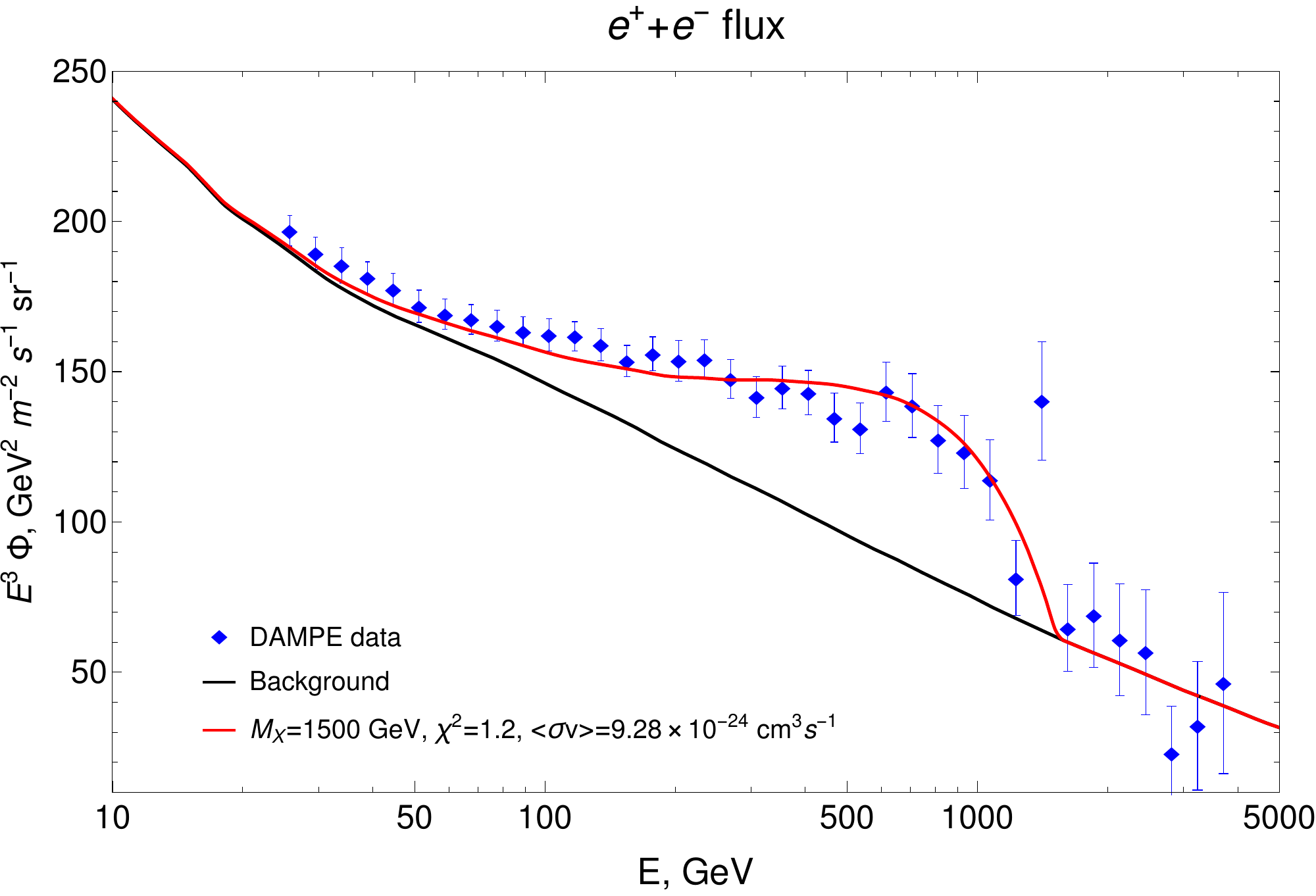}
    \includegraphics[width=0.5\textwidth]{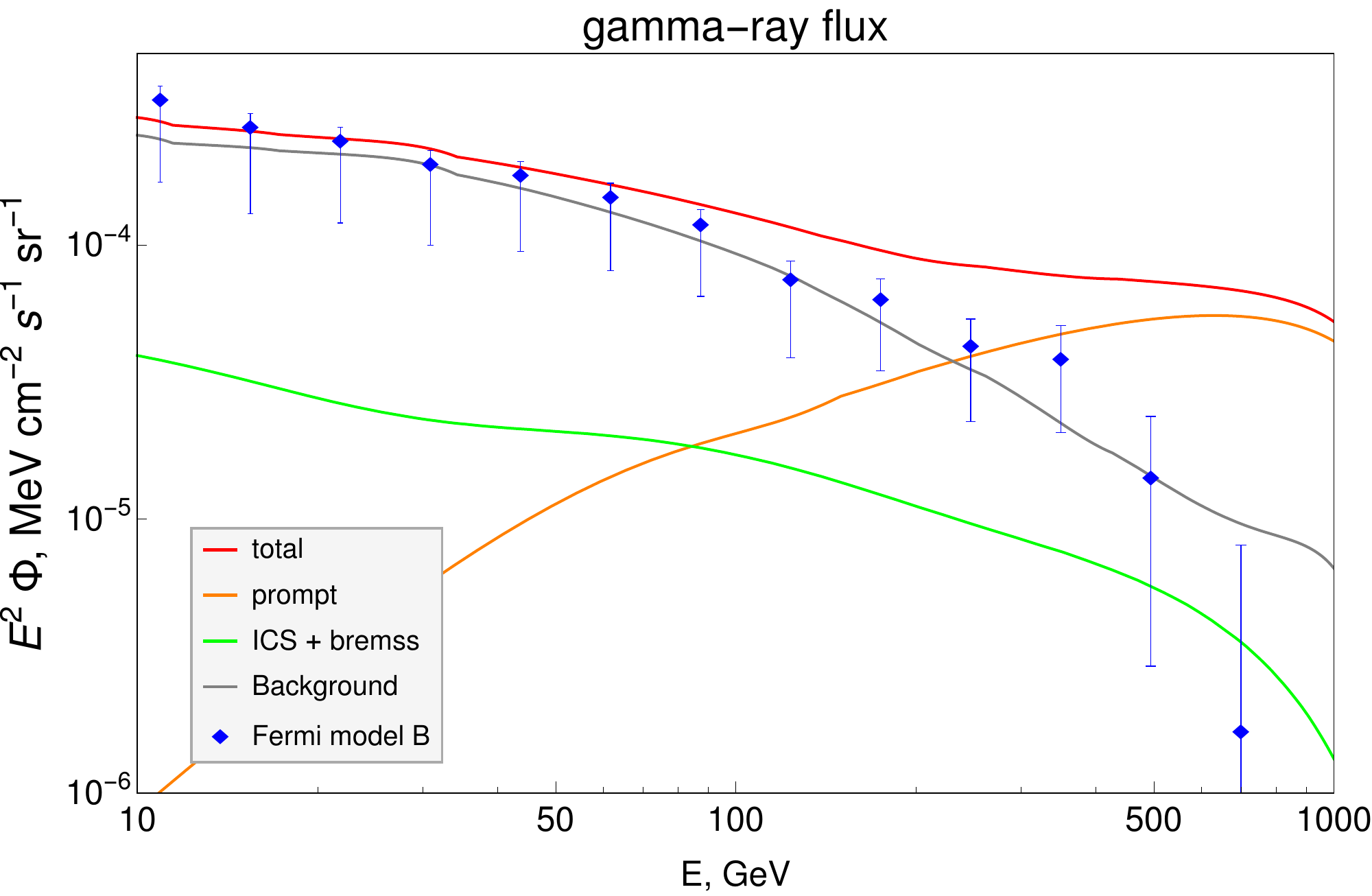}
    \label{4e_spectra_posi}}
    \subfigure[]{
    \includegraphics[width=0.49\textwidth]{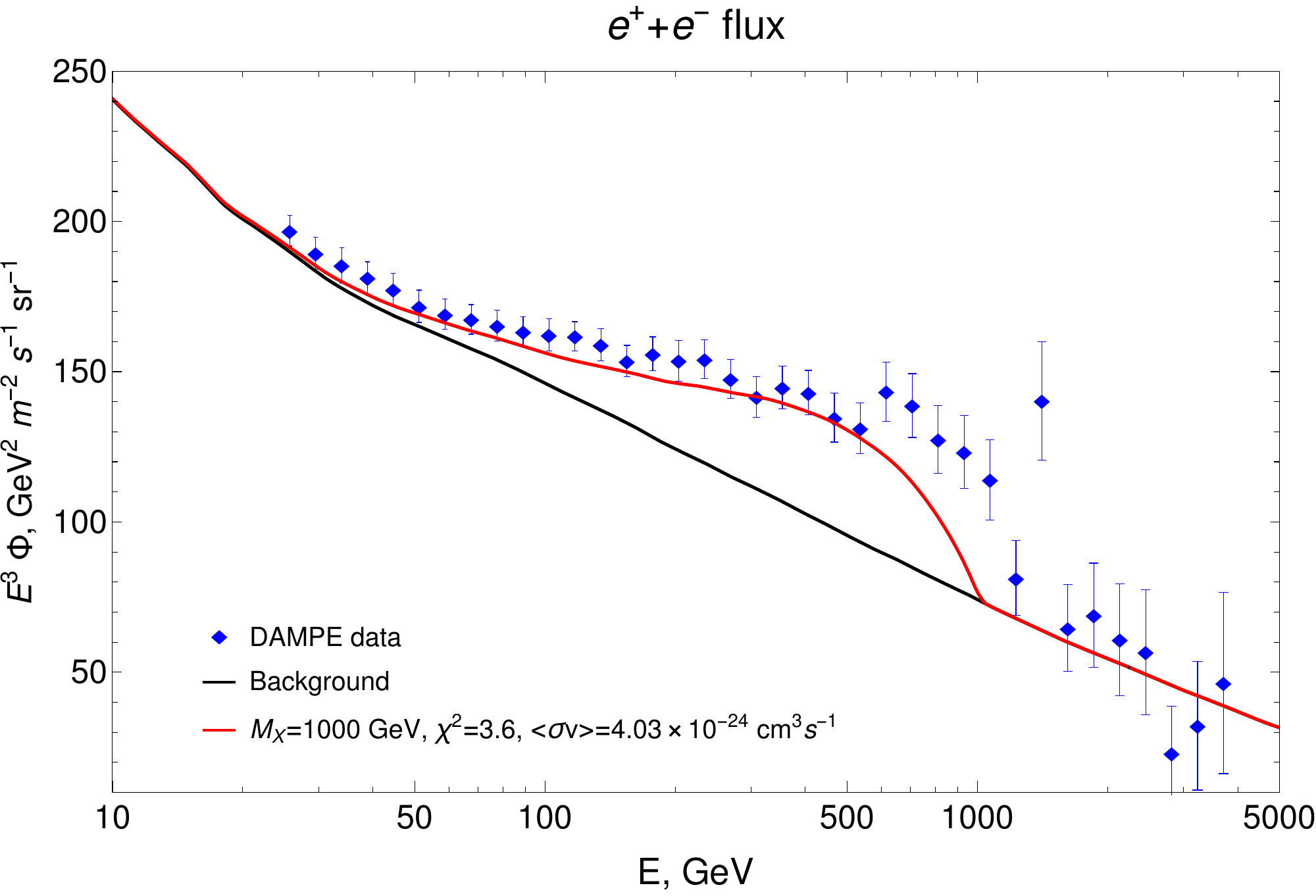}
    \includegraphics[width=0.5\textwidth]{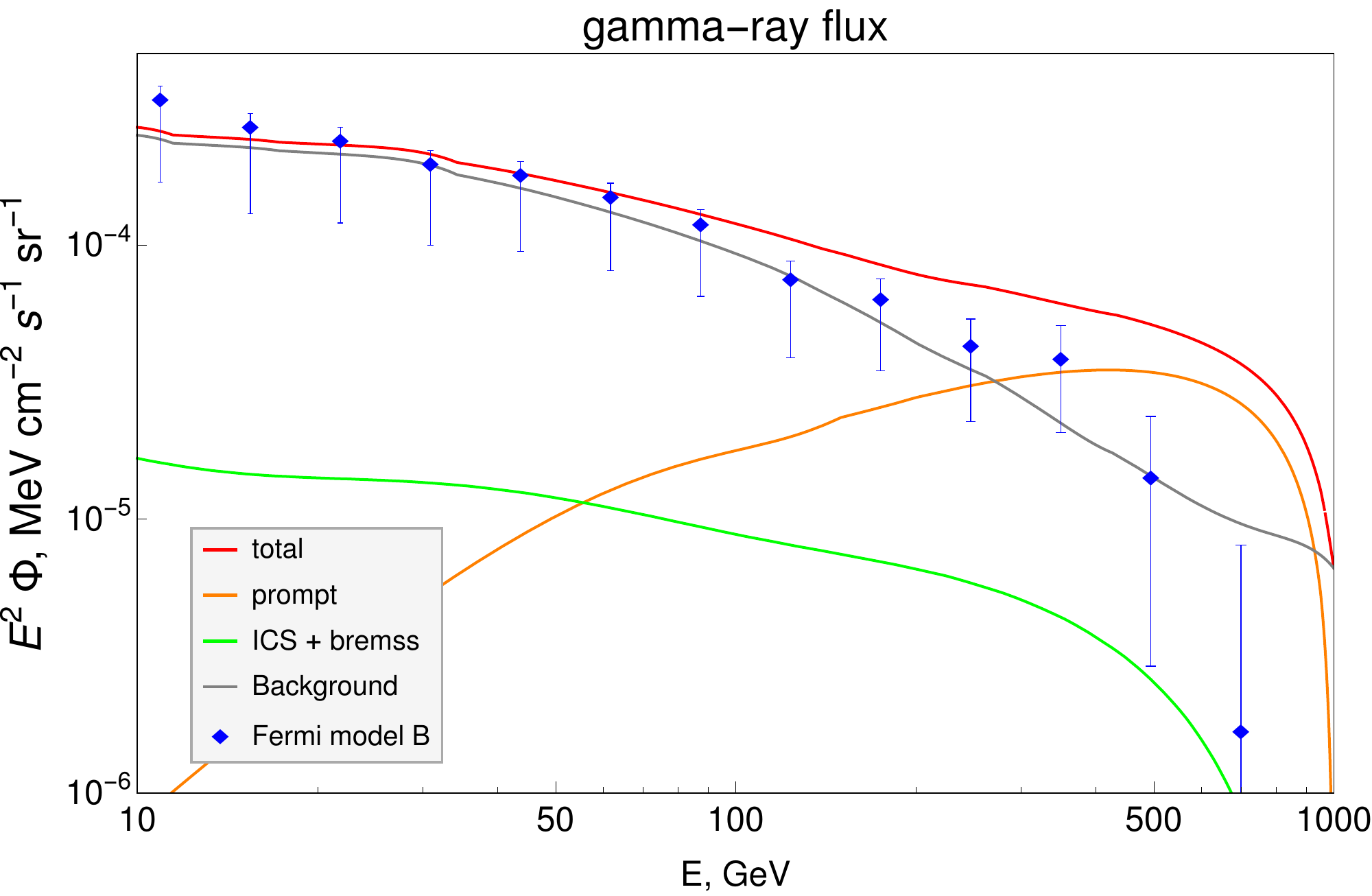}
    \label{4e_spectra_posigamma}}
    \caption{Same as Fig.~\ref{2e_spectra_extra}, but for the 1-step cascade model (Eq.~\eqref{X4}).}
    \label{4e_spectra_extra}
\end{figure}

\begin{figure}[p]
    \subfigure[]{
    \includegraphics[width=0.49\textwidth]{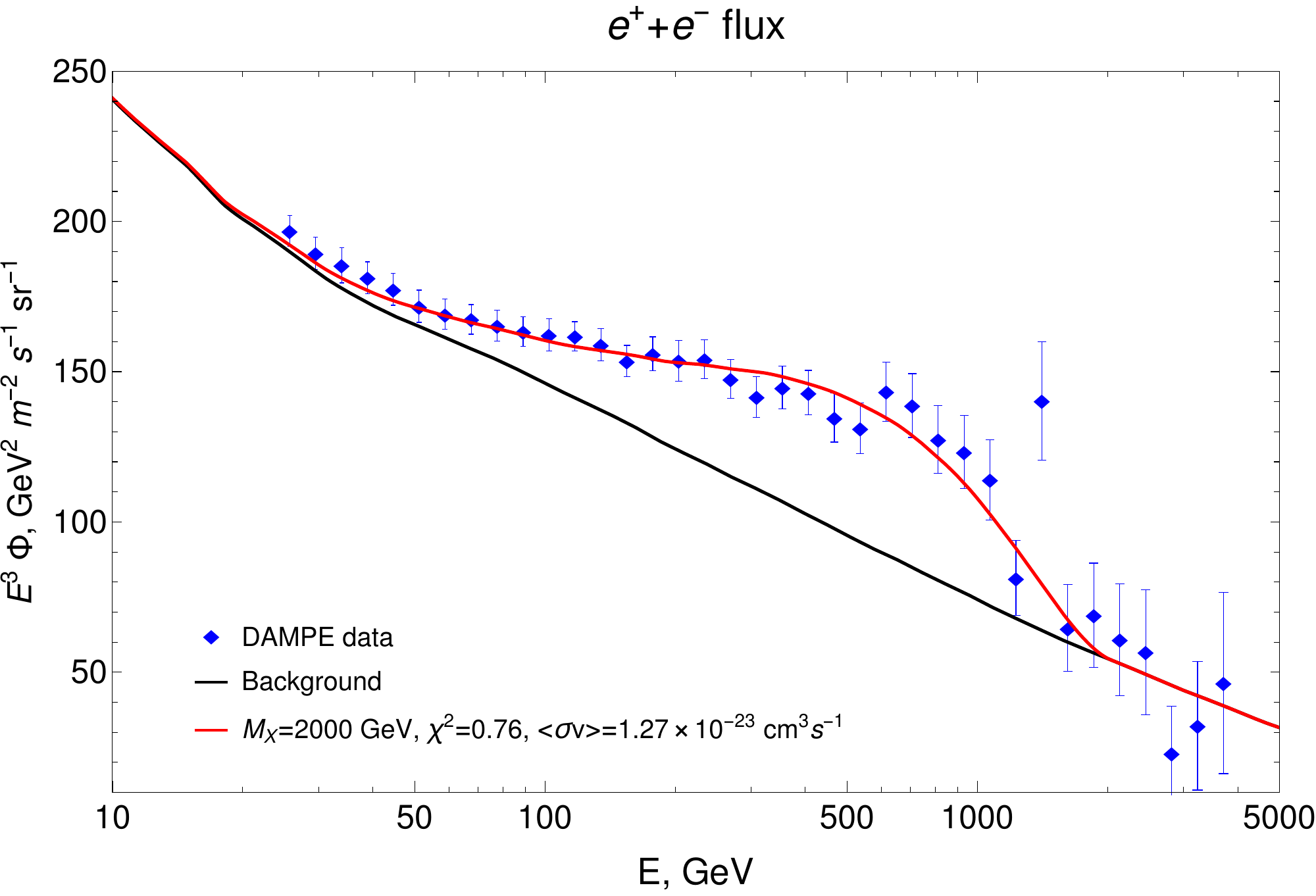}
    \includegraphics[width=0.5\textwidth]{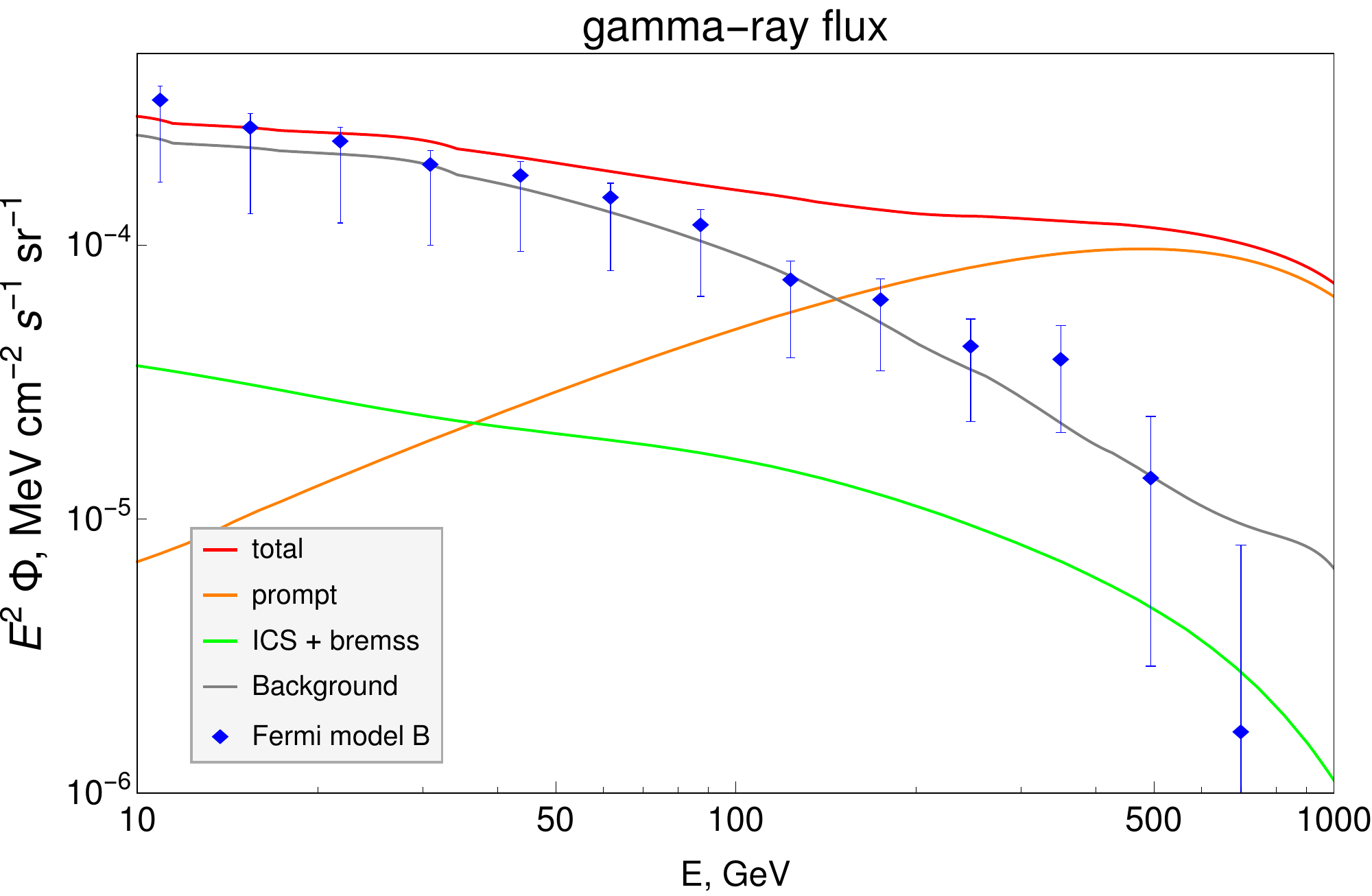}
    \label{8e_spectra_posi}}
    \subfigure[]{
    \includegraphics[width=0.49\textwidth]{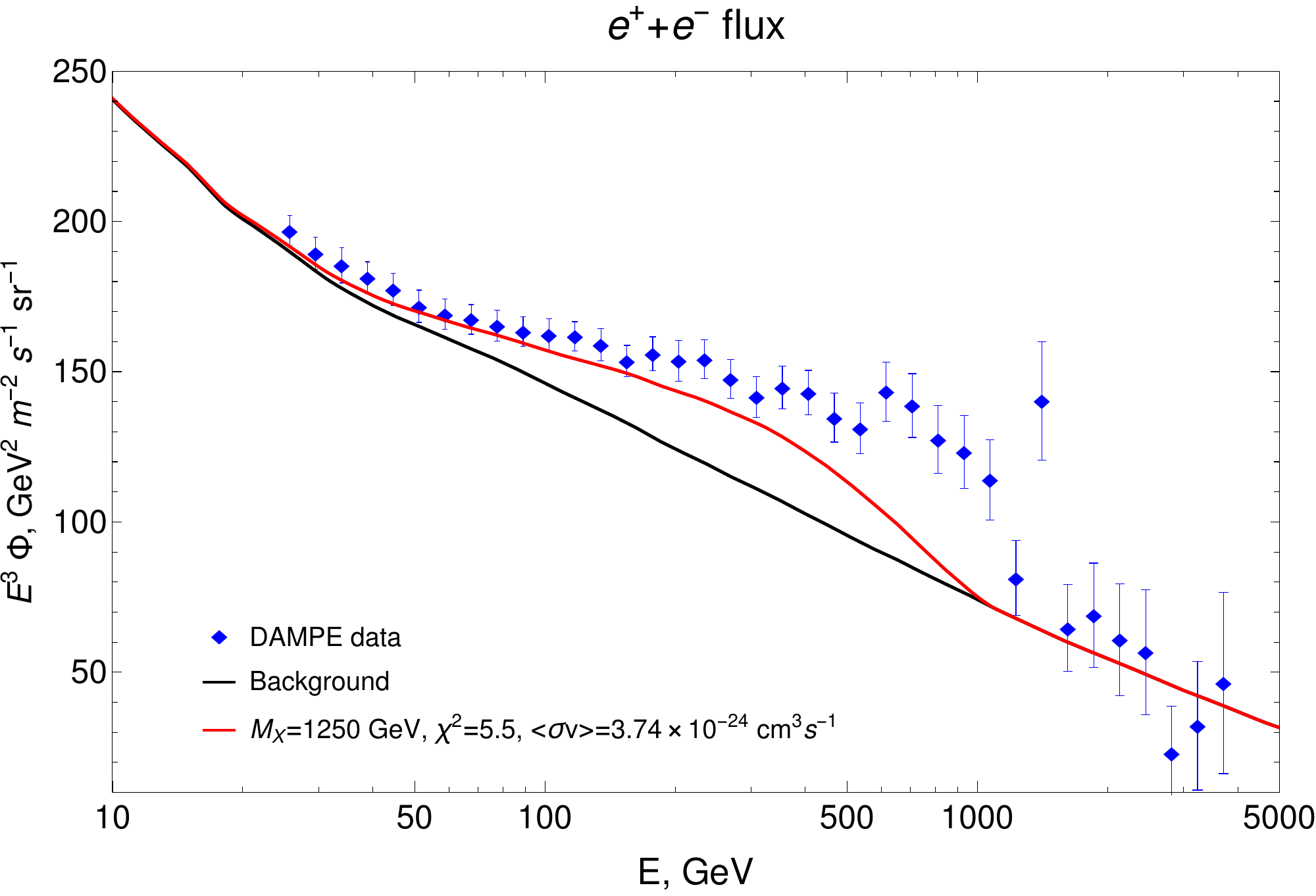}
    \includegraphics[width=0.5\textwidth]{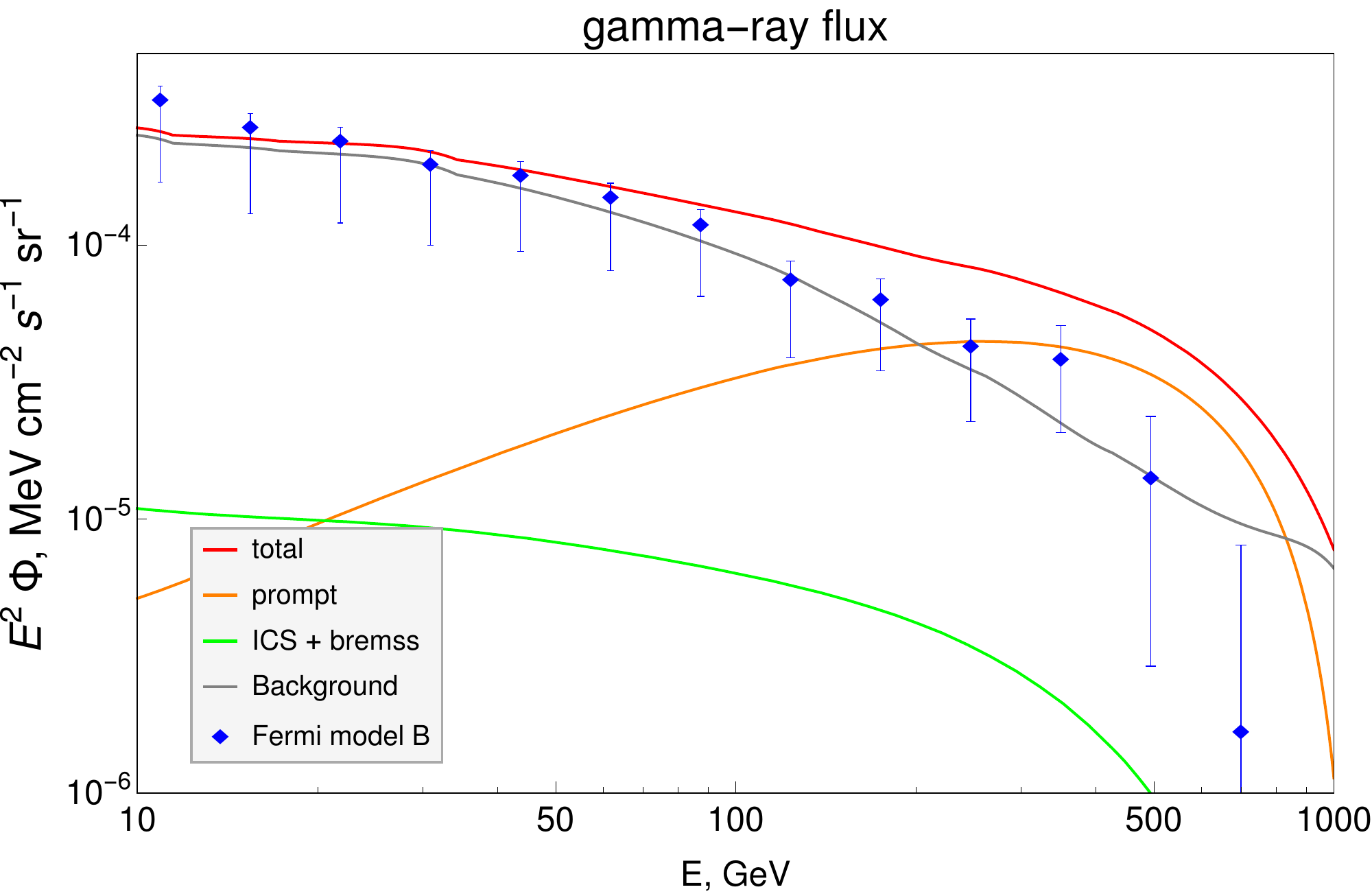}
    \label{8e_spectra_posigamma}}
    \caption{Same as Fig.~\ref{4e_spectra_extra}, but for
    2-step cascade model (Eq.~\eqref{X8}). Note, that the mass $M_X$ has to be increased comparing to the direct or 1-step cascade decays in order to fit the DAMPE data properly. }
    \label{8e_spectra_extra}
\end{figure}

\end{document}